\documentclass[prx,aps,twocolumn,showpacs,superscriptaddress,floatfix]{revtex4-1}
\usepackage{graphicx}
\usepackage{bm}

\usepackage[dvips]{color}
\usepackage{graphicx}

\begin{document}

\title{Fermionic spinon and holon statistics in the pyrochlore quantum spin 
liquid}

\author{B. Normand}
\affiliation{Department of Physics, Renmin University of China, Beijing
100872, China}

\author{Z. Nussinov}
\affiliation{Department of Physics, Washington University, St. Louis,
MO 63160, U.S.A.}

\date{\today}

\begin{abstract}
The one-band Hubbard model on the pyrochlore lattice contains an extended 
quantum spin-liquid phase formed from the manifold of singlet dimer coverings. 
We demonstrate that the massive and deconfined spinon excitations of this 
system have fermionic statistics. Holonic quasiparticles introduced by doping 
are also fermions and we explain this counterintuitive but general result. 
\end{abstract}


\maketitle

\section{Introduction}

Quantum spin liquids (QSLs) have become a focus of intense activity
in theory, numerics, experiment, and materials growth. Theoretical 
interest is driven by the possibility of understanding unconventional 
gapped and gapless quantum ground states, including their entanglement, 
topological properties, and fractional elementary excitations \cite{rb}. 
Although many QSL properties have been studied by considering somewhat 
abstract quantum dimer models (QDMs) \cite{rrk,rms}, Kitaev models \cite{rak}, 
SU(N) \cite{rhsflnw}, and other models, only recently have they been proven 
exactly in a physically relevant Hamiltonian, the one-band Hubbard model on 
the pyrochlore lattice \cite{rnn}. In common with more abstract QSLs, the 
pyrochlore QSL is based on a highly degenerate ground manifold (of 
nearest-neighbor dimer coverings), occurring at an exactly solvable Klein 
point in a frustrated spin model \cite{rbt,rnbnt}. An exact treatment of 
perturbations about this point reveals an extended region of parameter space 
where the ground state is a three-dimensional (3D) QSL with massive and 
deconfined spinon excitations.

A fundamental question about any QSL concerns the statistics of its 
quasiparticles. The connection of the intrinsic spin to the statistical 
nature of a particle dates back to Pauli \cite{rwp}. In the absence of 
Lorentz invariance, as in a solid, and in the presence of strong interactions, 
new options exist for the statistics of ``emergent'' low-energy quasiparticles. 
The best-known examples are the quasiparticles of the fractional quantum Hall 
effect \cite{rfqhe1,rfqhe2,rfqhe3}, which have fractional (or anyonic) 
statistics; similar effects have been sought in high-temperature 
superconductors \cite{rhtsc1,rhtsc2} and other models \cite{rfb}, 
including (chiral) QSLs \cite{rkl,rwwz,rzgs} and quantum critical systems 
\cite{rdqcp,rs}. Although not necessarily fractional, quasiparticles in 
these models may nevertheless contradict the spin-statistics theorem, such 
as the bosonic $S = 1/2$ spinons discussed in Refs.~\cite{rw,rss}.

Here we investigate the quasiparticle statistics of the pyrochlore QSL. 
Because all states of the ground manifold are known exactly, as are all 
transition matrix elements, this system may be understood completely and 
used to extend existing QSL knowledge. We compute the statistics of spinons, 
demonstrating that they are fermionic. We then find that holons, the charged 
quasiparticles obtained by doping the QSL, are also fermions. We demonstrate 
that this result has a simple electronic explanation and establish the 
connection of these emerging fermions with gauge fields, represented by 
strings, as anticipated in Ref.~\cite{rlw}.

\begin{figure}[t]
\includegraphics[width=6cm]{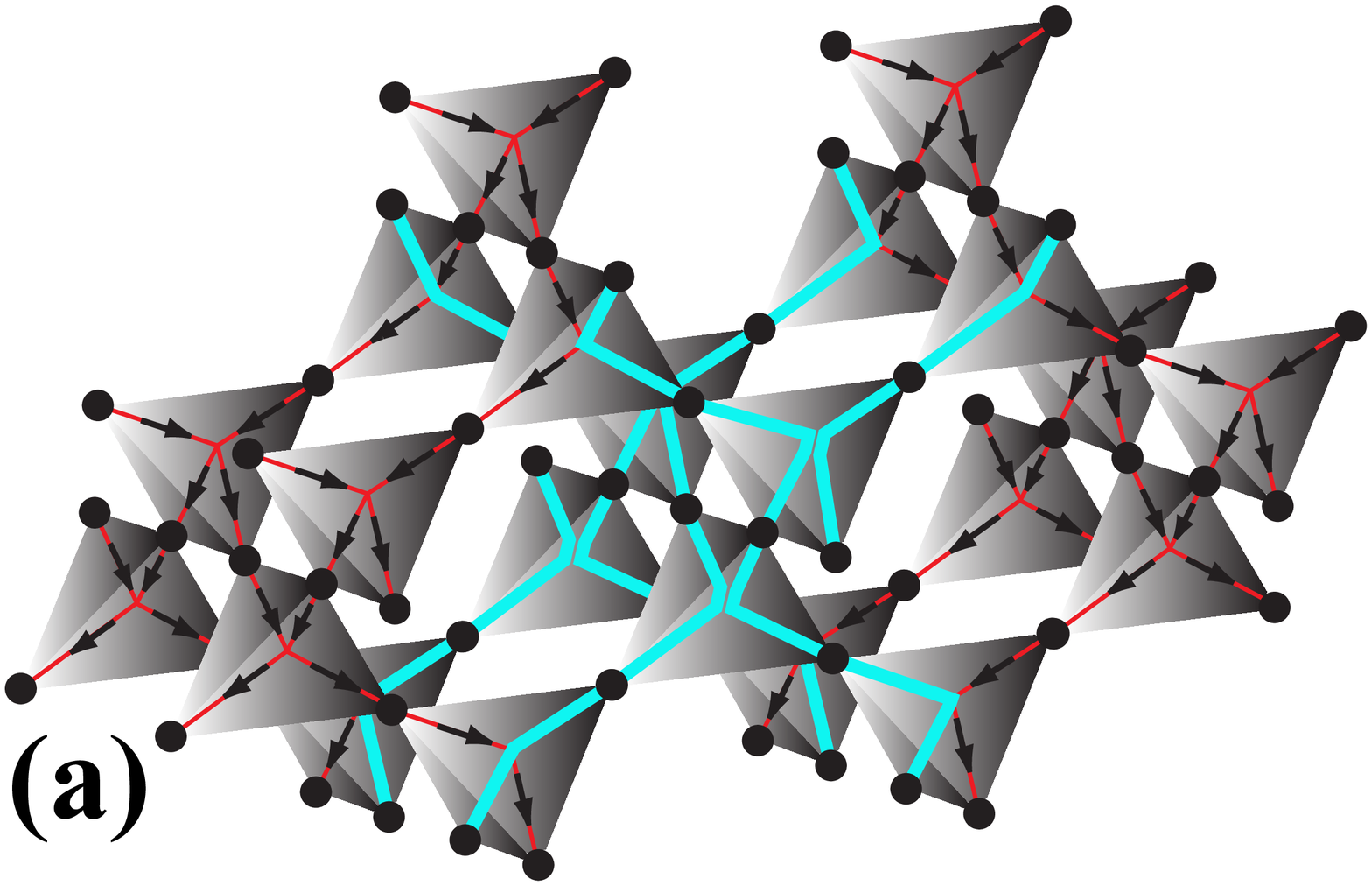}
\includegraphics[width=4.5cm]{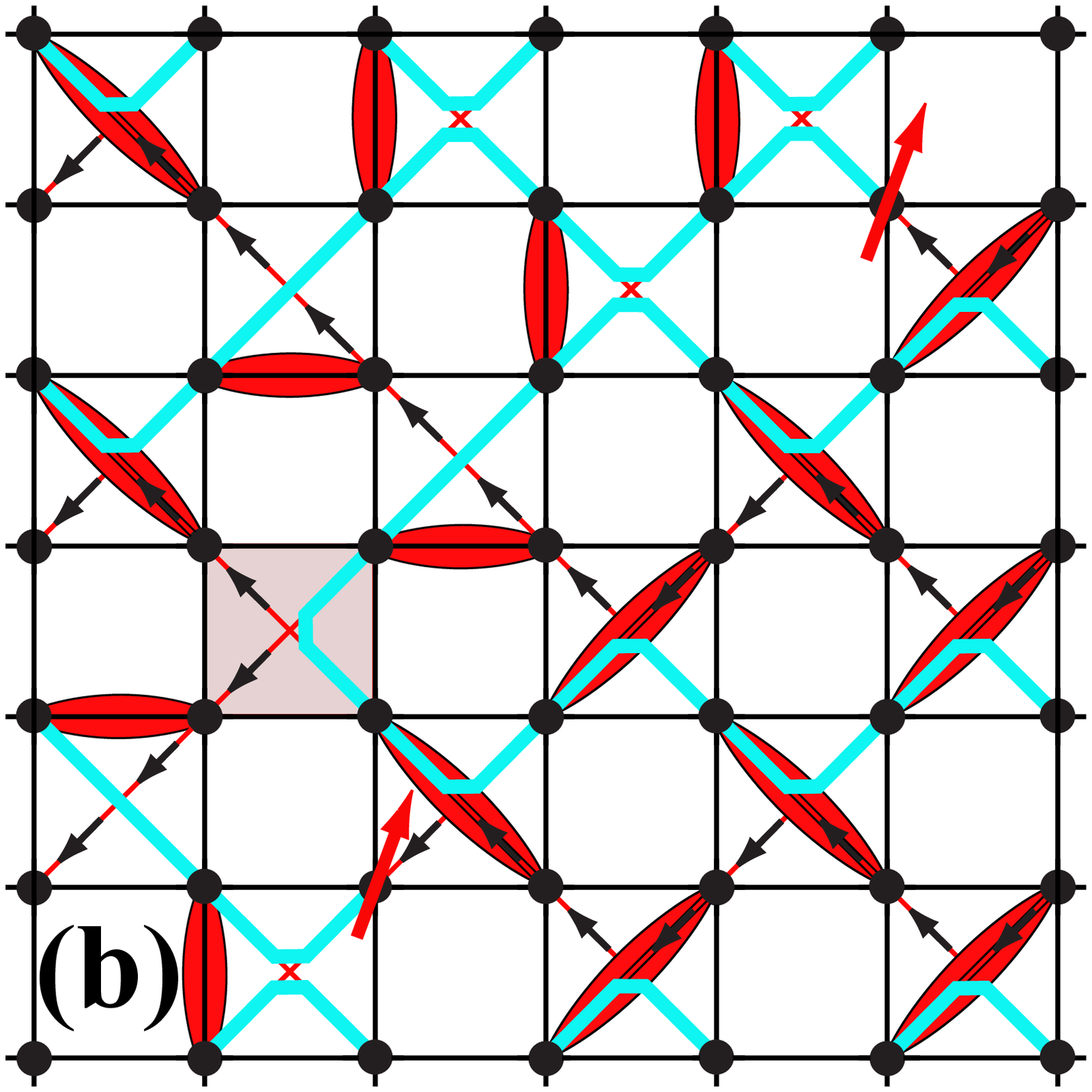}
\caption{(color online) (a) Pyrochlore lattice. The dimer covering (shown 
in Fig.~1 of Ref.~\cite{rnn}) is replaced both by its six-vertex representation 
(black arrows) and by the line representaion (light blue). Lines are drawn on 
all up-pointing arrows and do not cross in the ($xz$) plane. (b) Spinon motion 
represented on the checkerboard lattice, where lines are drawn on all 
right-pointing arrows and do not cross in ${\hat y}$ \cite{rnbnt}. Local 
(spinon-dimer) processes allow the spinons to move off the line changing 
direction at the DT (shaded plaquette).}
\label{fig1}
\end{figure}

Although we focus for exactness on the pyrochlore QSL, our considerations 
regarding the fermionic statistics of both spinons and holons are ubiquitous. 
We will show that they are related directly to the fundamental underlying 
statistics of the constituent electrons. Thus our results are universal for 
degenerate dimer-based electronic systems, including resonating valence-bond 
(RVB) states, and are by no means specific only to one model. 

The structure of this article is as follows. In Sec.~II we review the nature 
and known properties of the pyrochlore QSL. In Sec.~III we discuss how best 
to calculate the statistics of spinon excitations in the pyrochlore geometry
and use this method to demonstrate that spinons are fermions. In Sec.~IV we 
introduce the doped pyrochlore QSL and show that the resulting holon 
quasiparticles are also fermions; we explain this inobvious result and 
demonstrate its generality. In Sec.~V we implement a lattice gauge-theory 
representation of the local conservation law of dimer number to extend our 
analysis of emerging fermions by including their U(1) gauge content and its 
representation as strings. In Sec.~VI we discuss the physical relevance of 
strings by comparison with the line representation and clarify the issue of 
the local nature of quasiparticles. Section VII contains a brief summary and 
conclusion.

\section{Pyrochlore Quantum Spin Liquid}

The pyrochlore lattice is a 3D array of corner-sharing tetrahedra [Fig.~1(a)].  
The low coordination number, prevalence of triangles, and relevance of the 
``ice rules'' \cite{rbf} all contribute a wealth of phenomena in frustrated 
magnetism, including those of semiclassical ``spin-ice'' materials \cite{rcms}. 
The existence and properties of the pyrochlore QSL arise largely from the 
four-site symmetry of the tetrahedra and the zero-divergence condition [two 
in- and two out-pointing arrows in the six-vertex representation of Fig.~1(a)] 
encoded in the ice rules. 

Here we outline the sequence of logic proving the existence of the 
pyrochlore QSL \cite{rnn}. 1) A physically realistic Hamiltonian, the 
Hubbard model at half filling with pyrochlore geometry, gives a spin model 
with very strong fourth-order contributions. 2) These place it close to a very 
high-symmetry point, a Klein point, where the Hamiltonian is intra-tetrahedal 
only and can be expressed as a sum of projectors. 3) All states with one dimer 
per tetrahedron are exact ground states of this model. 4) There is a one-to-one 
mapping from this manifold of states to the six-vertex model (which encodes the 
ice rules). 5) This ground manifold has extensive degeneracy, among whose 
consequences is that all states of the manifold are connected to other states 
by zero-dimensional (small-loop) processes. 6) The zero-divergence condition 
of the ice rules is a local conservation law, which results in the U(1) gauge 
nature of the system (Coulomb phase). 

7) By constructing the submanifold maximizing the number of local dimer 
fluctuation processes around hexagons (i.e.~involving six tetrahedra), which 
control the behavior under physical perturbations away from the Klein point, 
one may demonstrate the persistence of a highly degenerate ground manifold of 
${\cal O} (2^{L/3})$ basis states, where $L$ is the linear dimension of the 
system. Loop calculations for the physical processes linking states of the 
new ground manifold show that all such states gain energy from mutual 
resonance and that their linear combinations span all dimensions and break 
no lattice symmetries. 8) To complete the proof of a QSL ground state, one 
demonstrates that the states of the new ground manifold satisfy rigorous 
topological criteria, based on the presence of both local (0D) and emergent 
planar (2D) gauge-type symmetries, respectively of U(1) and Z$_2$ type. Thus 
one may conclude the existence of a true, zero-temperature, 3D QSL occurring 
over an extended region of the model parameter space around the Klein point.

Next we summarize the energetic properties of the spin excitations of 
the pyrochlore QSL, which arise as a necessary consequence of its nature. 
a) The destruction of one singlet creates a defect tetrahedron (DT) with no 
dimer. b) This process creates two individual spins in a net triplet state 
and comes at an energetic cost \cite{rnn}, meaning that the spin excitations 
are massive. c) Once created, the two individual spins may propagate 
separately at no further energetic cost by the rearrangement of dimers; 
hence it is appropriate to regard these $S = 1/2$ objects as spinons, which 
further are deconfined, their quantum dynamics allowing free propagation at 
$T = 0$. For a graphical representation of spinon motion, Fig.~1(a) 
illustrates both the six-vertex and line representations \cite{rnbnt} of a 
dimer configuration in the ground manifold of the pyrochlore QSL \cite{rnn}. 
For ease of visualization, we switch in Fig.~1(b) to the checkerboard (2D 
pyrochlore) lattice to illustrate the dimers, spinons, DTs, and their 
associated line representation \cite{cpc}. 

d) Off the Klein point, free spinon propagation is constrained by 
``dimensional reduction:'' the fact that the dimension of the ground 
manifold scales exponentially with $L$ (and not $L^3$) makes the system 
effectively 2D and the spinons move normal to its fluctuating planar degrees 
of freedom. This situation is encapsulated by the line representation of 
Fig.~1(a), where the planes are horizontal and the non-Klein-point ground 
states have on average of one line per tetrahedron (which is the maximally 
degenerate submanifold of the Klein-point states). e) In addition to planar 
fluctuations, local processes allow free movement of spinons from one line 
to another, as represented on the checkerboard lattice in Fig.~1(b). As a 
result, their motion is fully 3D and the situation is quite different from 
the strict linear spinon motion of Ref.~\cite{rbt}. f) The connection of 
the U(1) gauge field to the quasiparticles of the pyrochlore QSL has not 
yet been explored and is one subject of the current study; in fact we will 
demonstrate that it is the quasiparticles which make the U(1) field manifest. 

A further property of the pyrochlore QSL not discussed in Ref.~\cite{rnn} is 
the following. Gapped spinons, meaning with a singlet-triplet gap, mediate 
short-ranged spin-spin correlation functions. However, the ground manifold, 
both at and off the Klein point, is a highly degenerate set of singlet 
states. In this singlet manifold, dimer-dimer correlations are algebraic, 
i.e.~long-ranged, as shown explicitly in Ref.~\cite{rnbnt}. This type of 
state was classified in Ref.~\cite{rn} as a ``Type-II Gapped'' QSL and it 
arises in the pyrochlore Hubbard model as a consequence of the rigorous 
dimerization of all spin degrees of freedom (perfect singlet formation). 
Any system in which this process is only approximate could not display the 
two contrasting paradigms for QSL nature simultaneously. However, the 
definition of the pyrochlore QSL as ``gapped'' should be regarded as 
semantic only, or at best probe-dependent, because it may equally be 
classified as an algebraic dimer liquid.

\section{Spinon Statistics Calculations}

The statistical nature of quasiparticle excitations is not only an important 
characteristic of any strongly correlated quantum system but a fundamental 
question intrinsic to our understanding of the fabric of physics, namely the 
defining properties of bosons, fermions, and anything in between. From the 
standpoint of strongly interacting systems, studies of cuprate-inspired models 
\cite{rkrs,rk,rrc} and doped QDMs \cite{rlrcpp,rlropp} reveal that statistics 
are optional in 2D because flux attachment can interconvert bosons and 
fermions. In QDMs and also our checkerboard QSL, these are the only options, 
i.e.~anyons are excluded. 

Quasiparticle statistics are expected to be robust in 3D, but more challenging 
to compute because there is no unique exchange path. A definition of statistics 
based on the quasiparticle hopping algebra \cite{rlw} was recently specialized 
\cite{riqf} to compute the statistics of monomers in a 3D QDM from the 
expression $\theta_{ex} = \theta_s + \phi$. Here $\theta_{ex}$ is the flux 
(effective statistical angle) for exchanging two quasiparticles and $\phi$, 
the flux due to changes required of the dimer background to restore the 
initial state, is computed by taking a single quasiparticle around exactly 
the same path. $\theta_s$ is the ``true'' statistical angle intrinsic to the 
quasiparticles. 

We apply this procedure \cite{riqf} to compute the statistics of $S = 1/2$ 
spinons moving in the SU(2) dimer background of the pyrochlore QSL. The 
available local moves are obtained by applying the spin Hamiltonian 
\begin{equation}
H_t = \mbox{$\sum_l$} \, [ {\textstyle \frac12} J_1 {\bf S}_{l, \rm tot}^2 
+ {\textstyle \frac{1}{4}} J_2 {\bf S}_{l, \rm tot}^4 ],
\label{eht}
\end{equation}
which acts on each individual tetrahedron, $l$, whose total (four-site) spin 
is ${\bf S}_{l, \rm tot}$ \cite{rnn}. We will show below that, if a spinon is 
present on tetrahedron $l$, the effect of $H_t$ is to exchange the spinon 
position with one end of the dimer on the tetrahedron. The primary difficulty 
in the pyrochlore geometry is finding valid paths, composed only of these 
moves, which both exchange particles and restore the initial state. This 
requires additional moves only of the dimer background \cite{riqf}, which 
can be composed of the the minimal 8-(12-)bond loops in 2D (3D) systems, 
knwon as Rokhsar-Kivelson (RK) processes \cite{rrk,rnn}.

\begin{figure}[t]
\includegraphics[width=2.5cm]{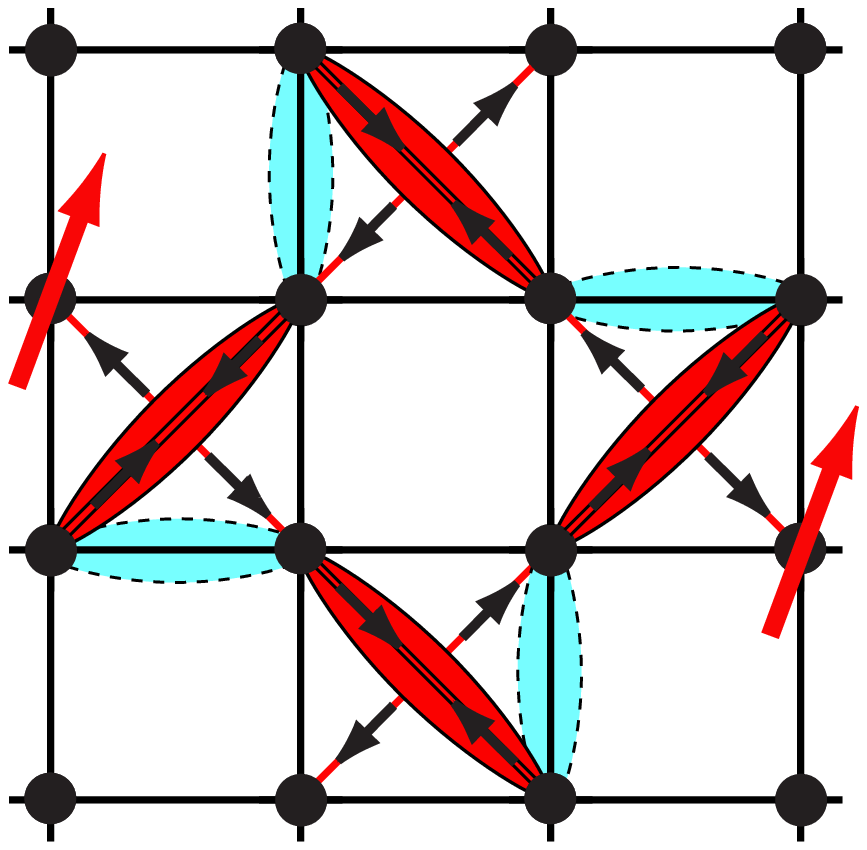}\hspace{0.6cm}\includegraphics[width=3.9cm]{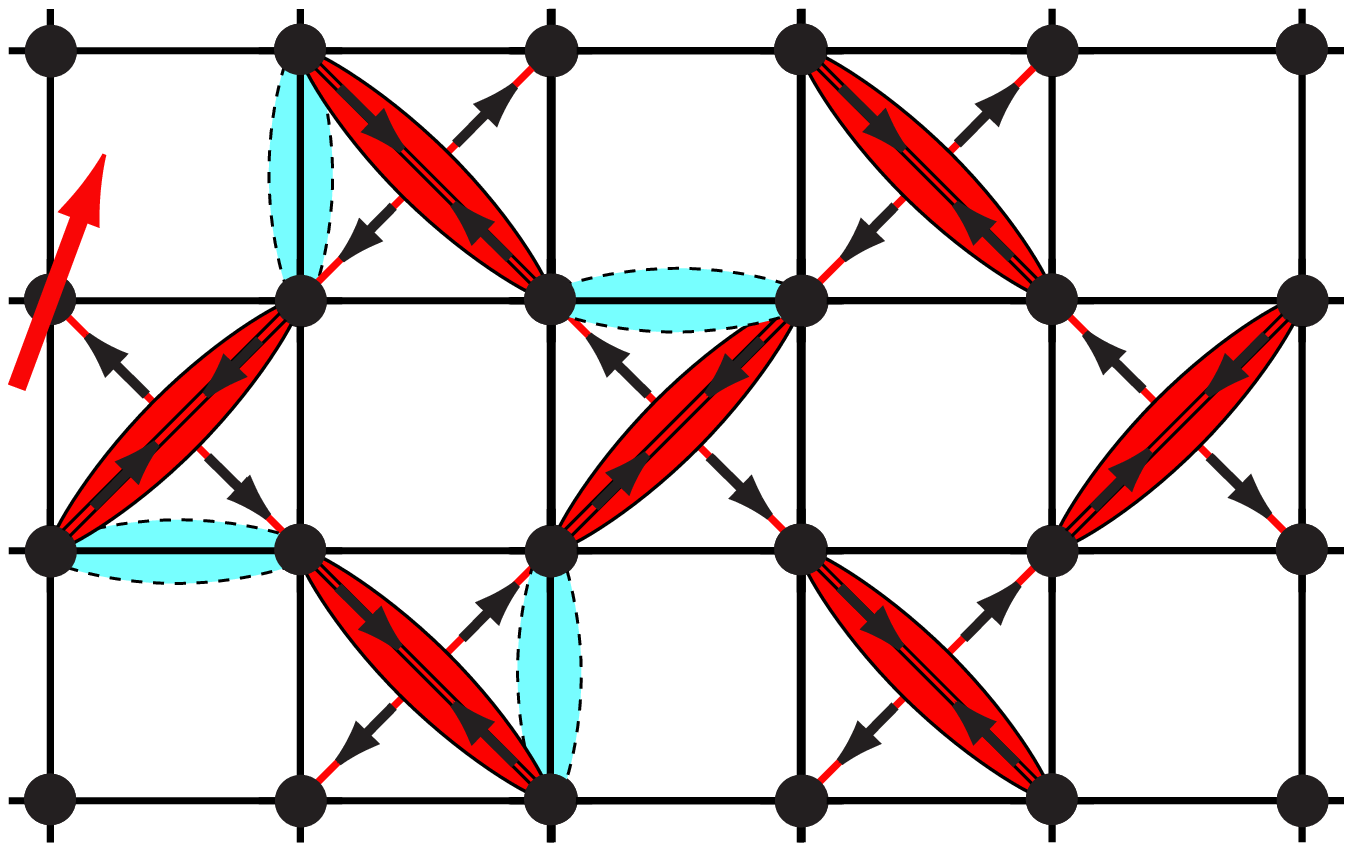}
\includegraphics[width=2.5cm]{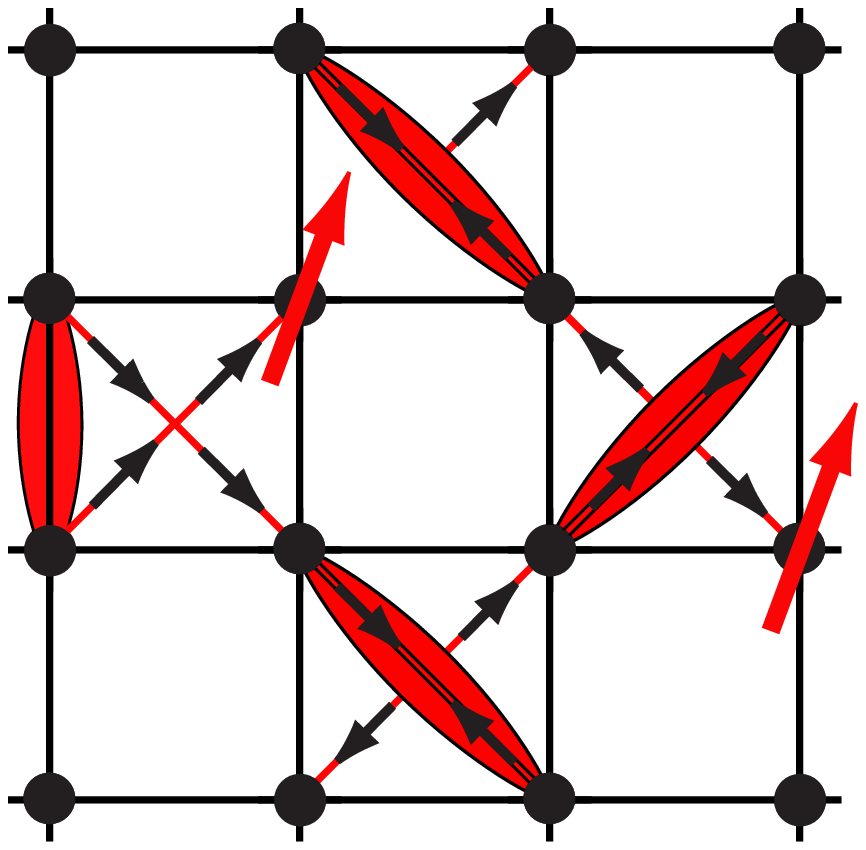}\hspace{0.6cm}\includegraphics[width=3.9cm]{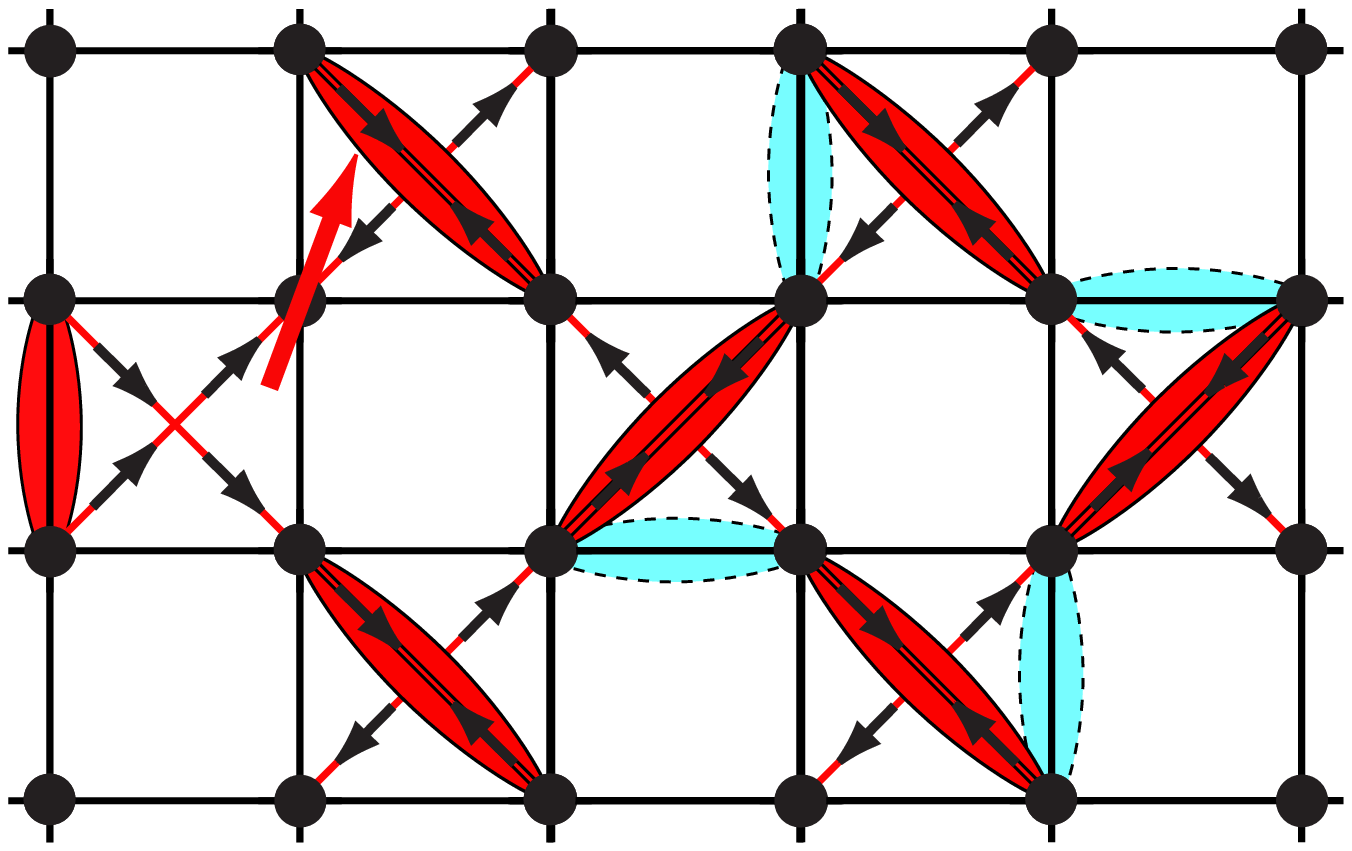}
\includegraphics[width=2.5cm]{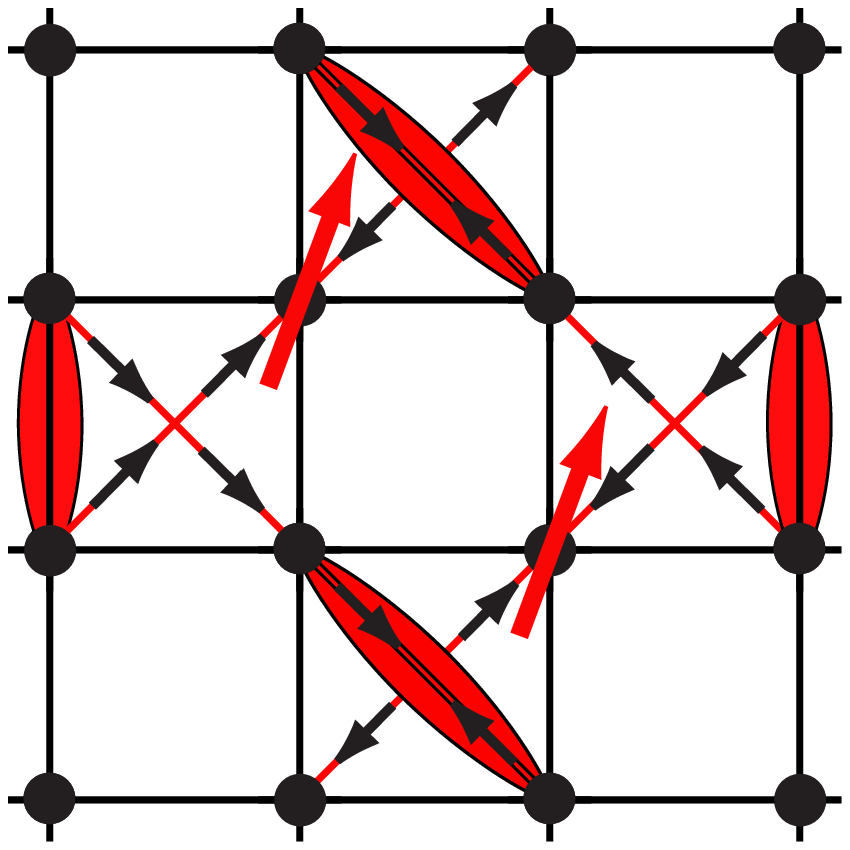}\hspace{0.6cm}\includegraphics[width=3.9cm]{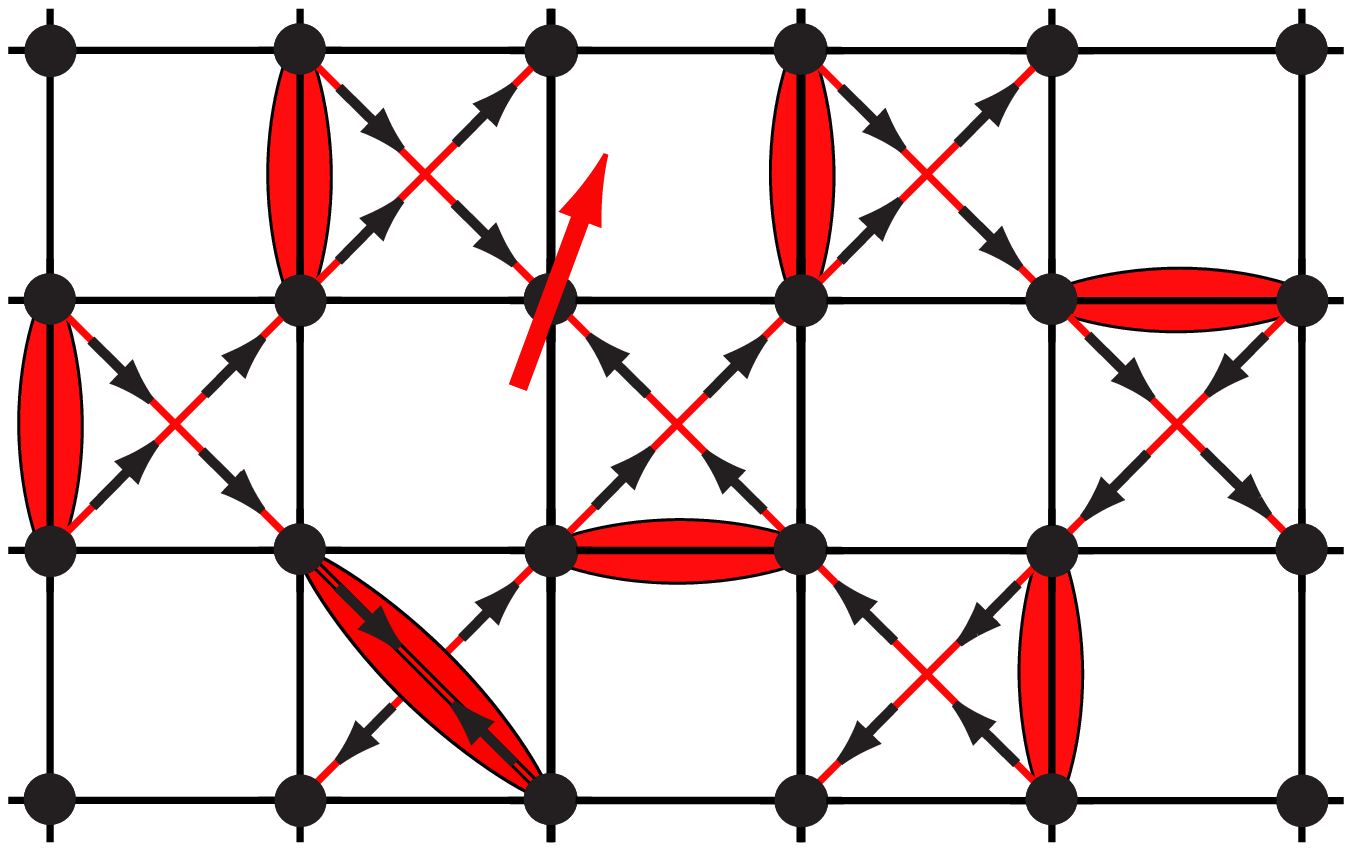}
\includegraphics[width=2.5cm]{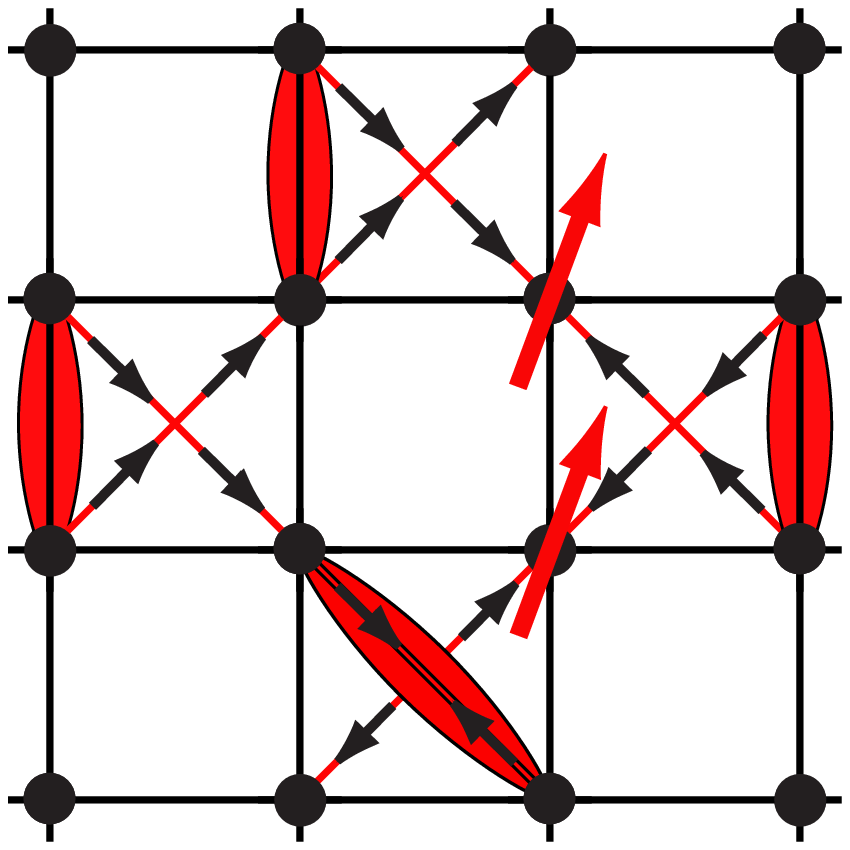}\hspace{0.6cm}\includegraphics[width=3.9cm]{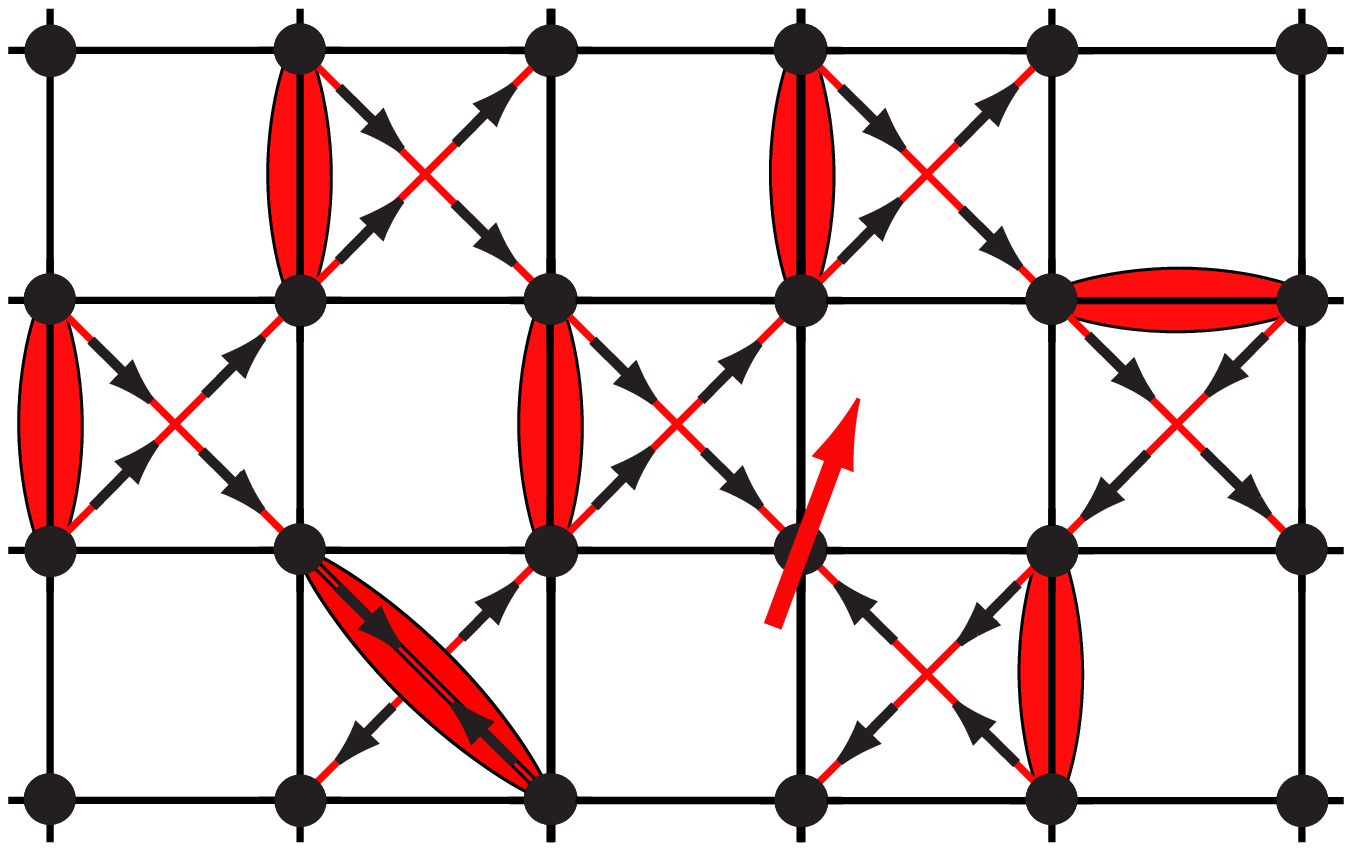}
\includegraphics[width=2.5cm]{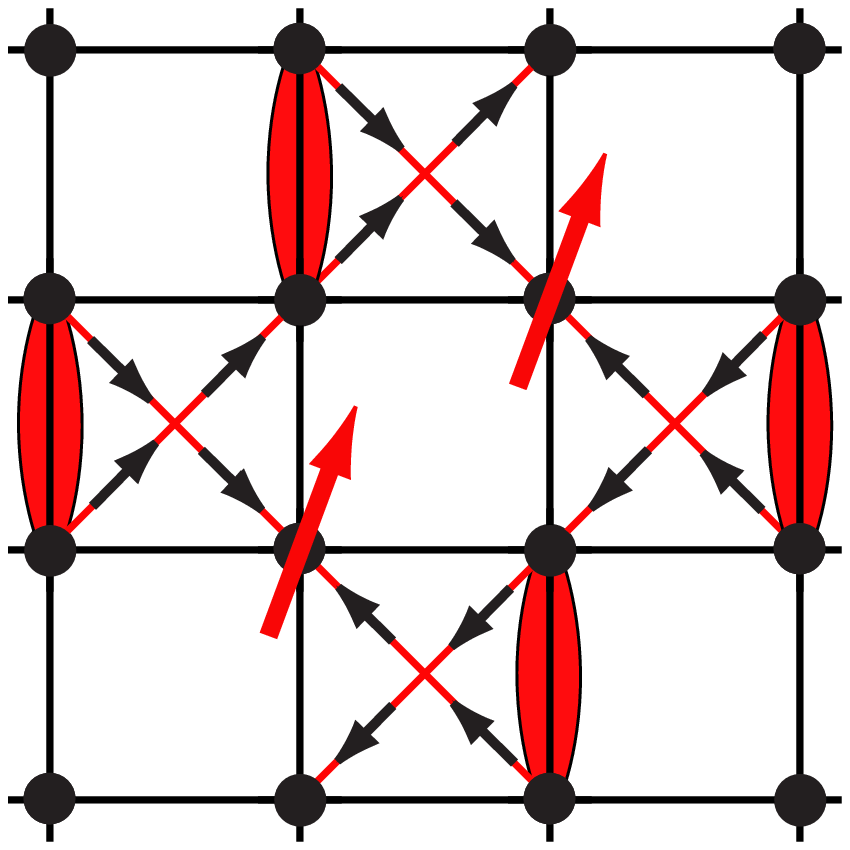}\hspace{0.6cm}\includegraphics[width=3.9cm]{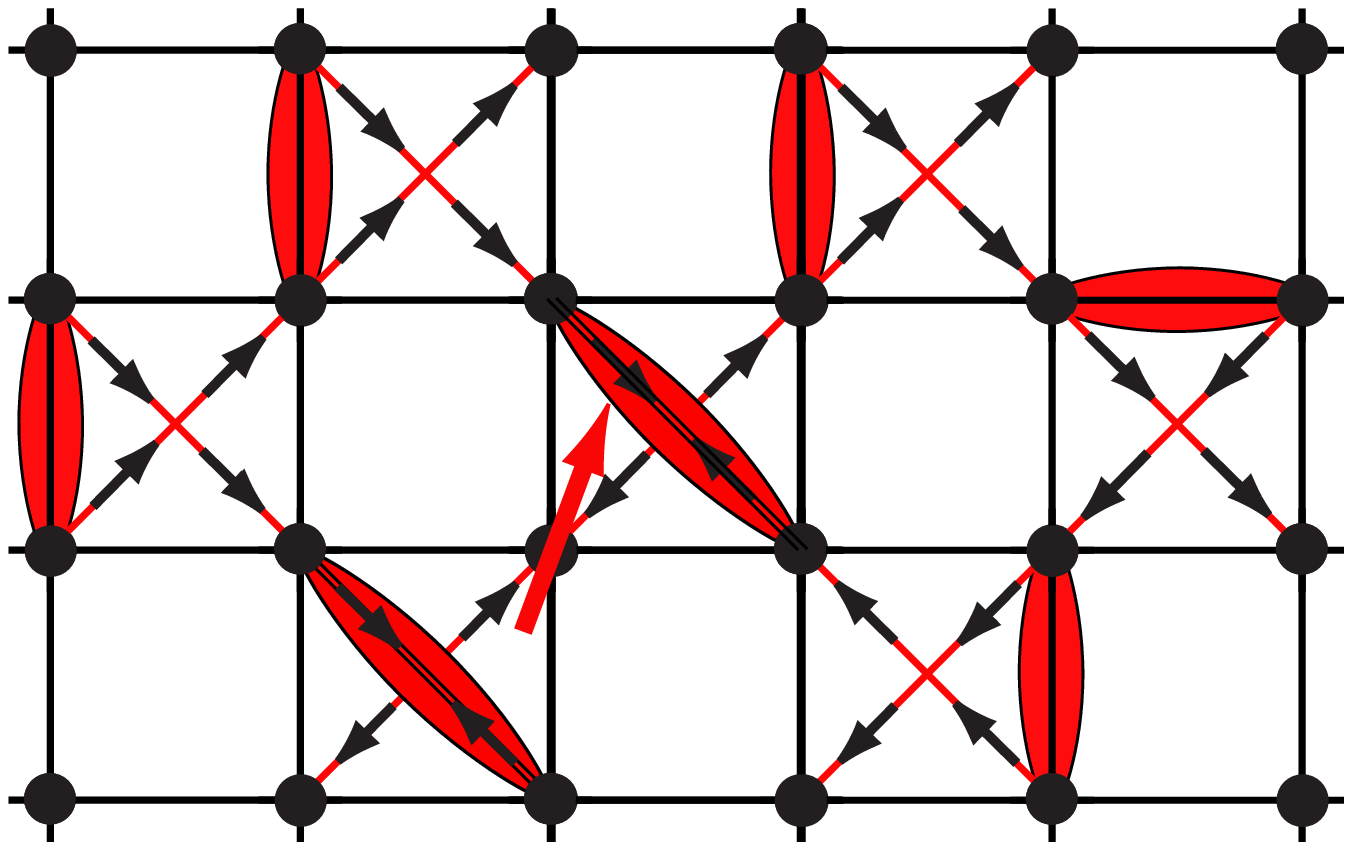}
\includegraphics[width=2.5cm]{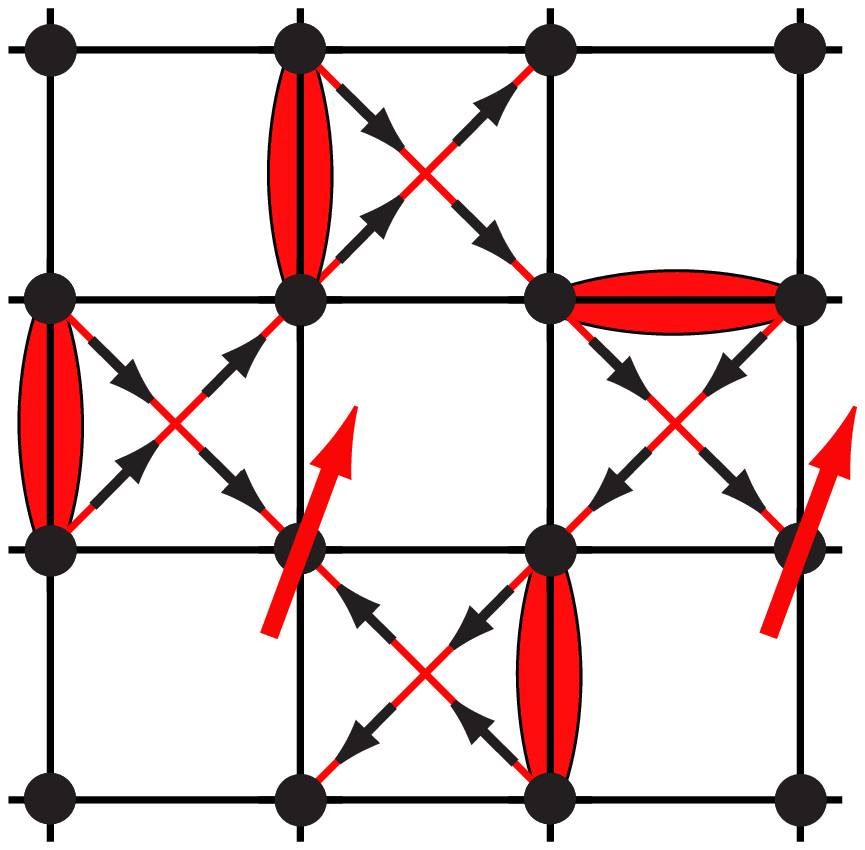}\hspace{0.6cm}\includegraphics[width=3.9cm]{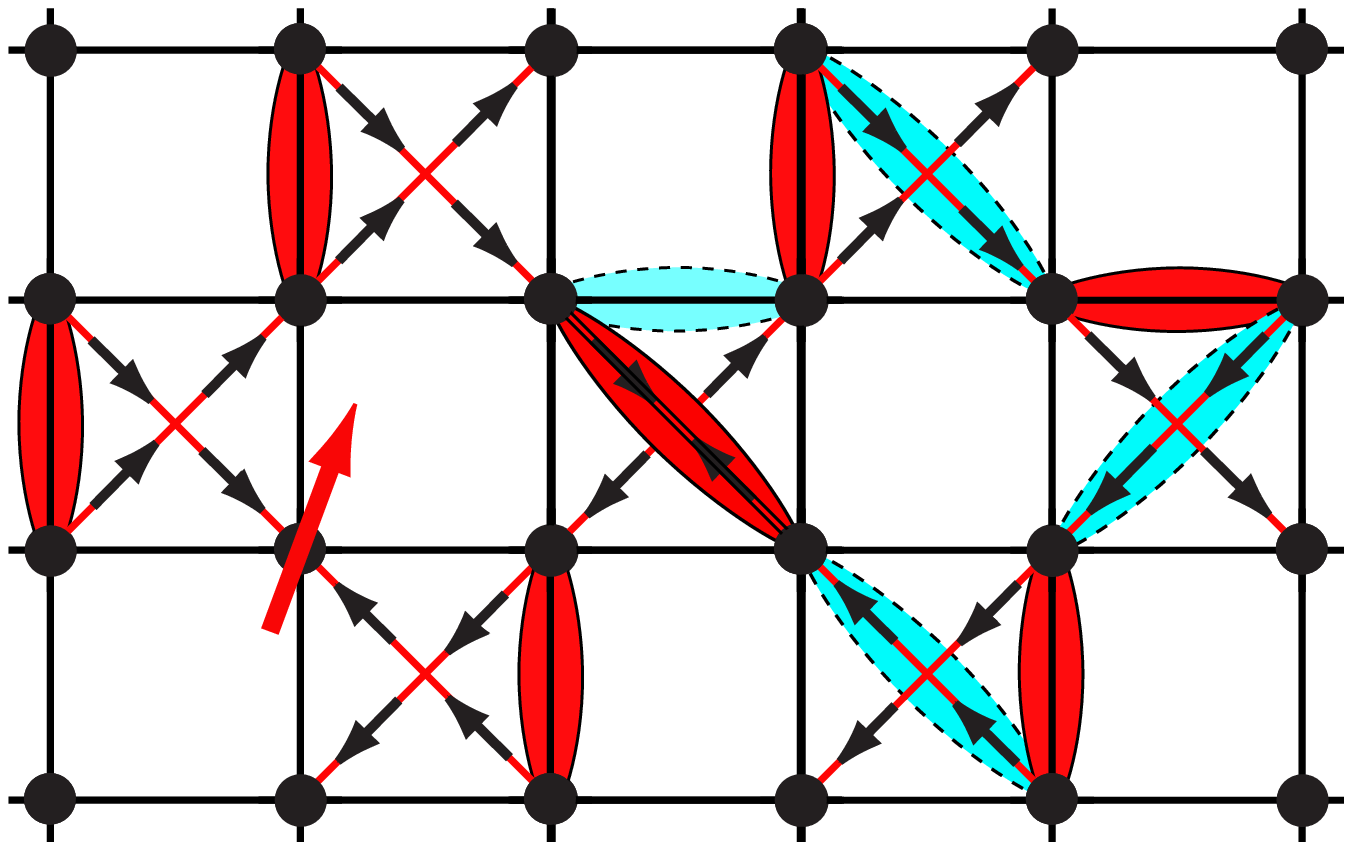}
\includegraphics[width=2.5cm]{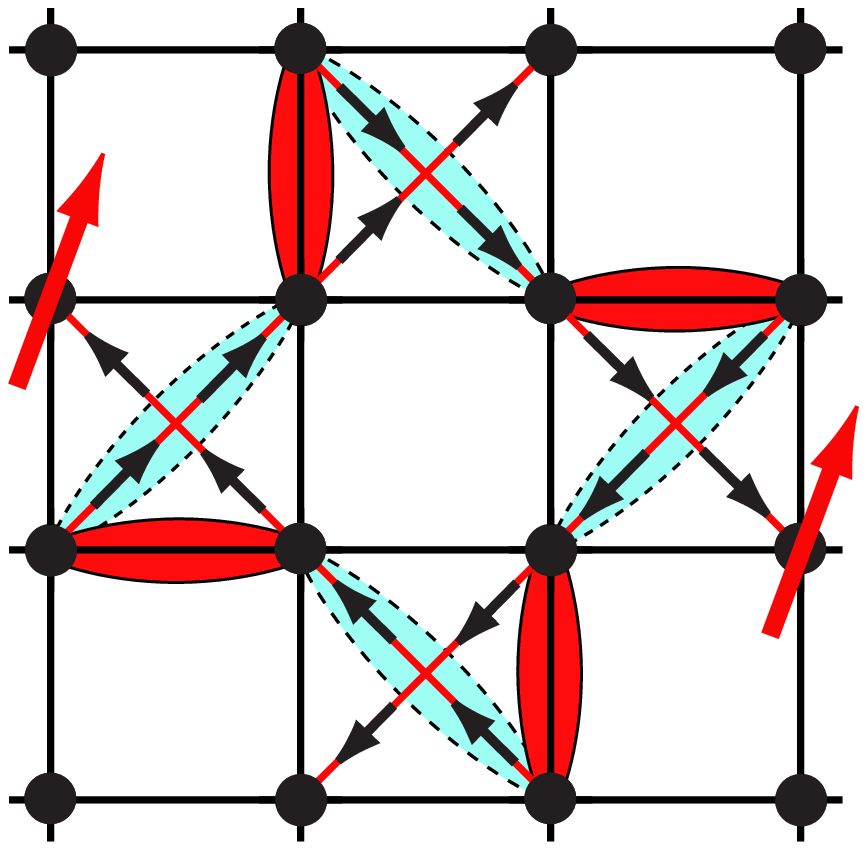}\hspace{0.6cm}\includegraphics[width=3.9cm]{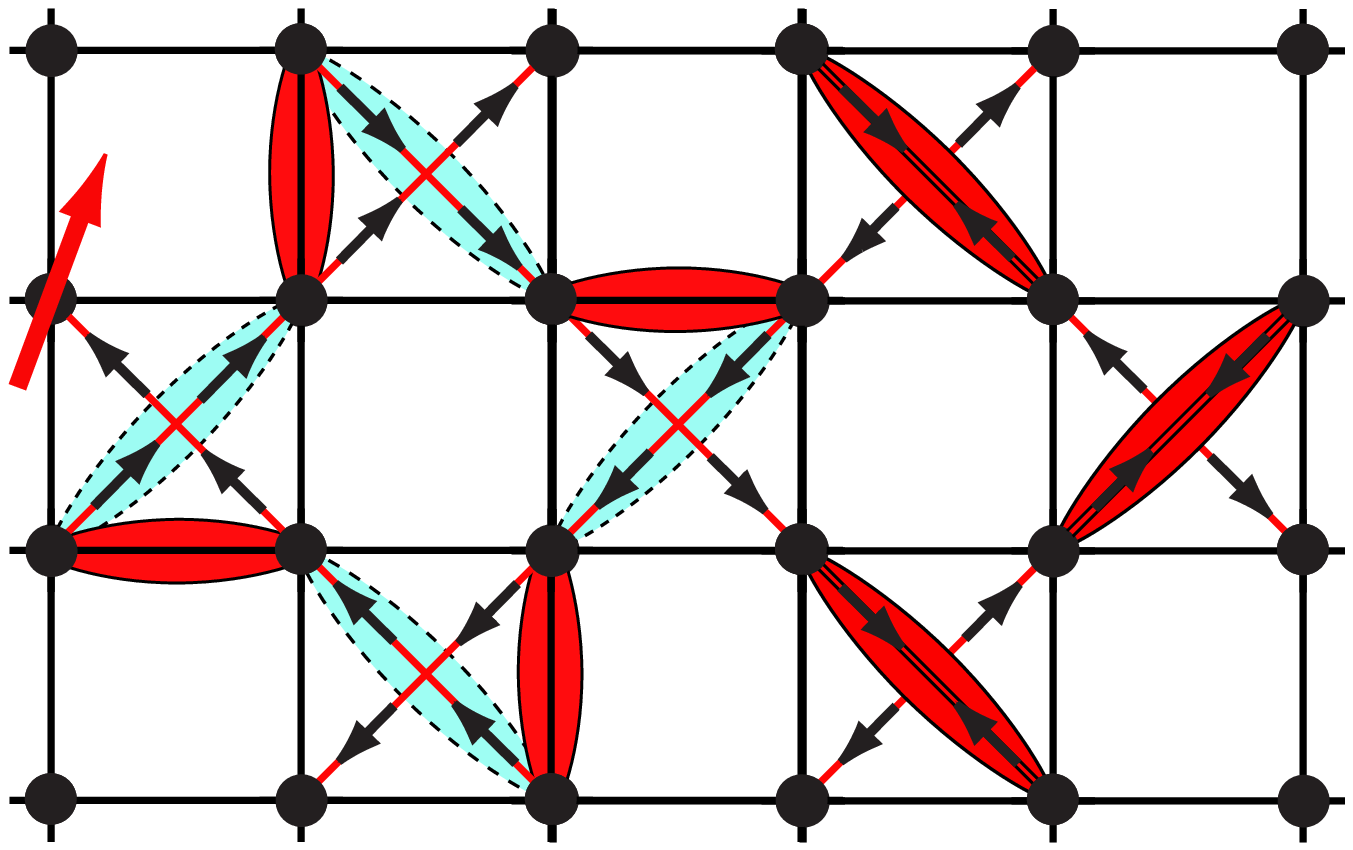}
\caption{(color online) Representation of spinon exchange processes on the 
checkerboard lattice. Left: the six steps required to exchange two spinons. 
Right: steps required for a single spinon to trace the same path. Red (gray) 
ellipses denote dimers and turquoise (light gray) ellipses the flipped dimer 
configuration resulting from an RK loop process.}
\label{fig2}
\end{figure}

One- and two-spinon processes valid for the considerations of exchange 
statistics are possible only when the quasiparticles are located next to, 
but not on, a dimer loop, as represented in 2D in the 1st and 7th panels of 
Fig.~2. Here, two-spinon exchange (left panels) alters the dimer background 
by one regularly-shaped RK loop and a single RK process is required to 
restore the initial state. When a single spinon moves on exactly the same 
loop (right panels), the initial dimer configuration blocks the site from 
which the second spinon began and the process requires a symmetrical RK loop 
(Fig.~2, 2nd right panel) acting in the neighboring square of the system. 
Restoring the initial state requires both an asymmetric RK process (6th 
panel) and the symmetric RK loop of the spinon path (1st and 7th panels). 

We compute the associated signs from a sign convention [Fig.~3(a)] with arrows 
oriented upwards ($+{\hat z}$) on a bond as the primary criterion and along 
$+{\hat x}$ as secondary. Loops are traversed clockwise. The sign contributed 
by each local process may be deduced only from the triangle of sites formed by 
the moving spinon and dimer. The relevant process in $H_t$, ${\vec S}_i.{\vec 
S}_k$, cast as the permutation operator $P_{ik} = 2 ({\vec S}_i . {\vec S}_k + 
1/4)$, acts on the three-spin state $|d_{ij} \rangle |\! \uparrow_k \rangle 
\equiv {\textstyle \frac{1}{\sqrt{2}}} (|\! \uparrow \downarrow \rangle - |\! 
\downarrow \uparrow \rangle) |\uparrow \rangle$ to produce the permuted state 
${\textstyle \frac{1}{\sqrt{2}}} (|\! \uparrow \downarrow \uparrow \rangle
 - |\! \uparrow \uparrow \downarrow \rangle) \equiv |\! \uparrow_i \rangle 
|d_{kj} \rangle$. This process changes the effective singlet sign on the loop, 
which traverses the sites in the order $ijk$; the same applies for a spinon of 
opposite spin, $|\downarrow_k \rangle$. The overall statistical angles are 
the products of these sign exchanges along the even-length loops of Fig.~2 
with the factors due to the RK loops. 

Because the symmetric RK loop contributes $+1$, the overall factor for 
two-particle exchange is $e^{i \theta_{ex}} = (-1)^6 \times (1)$, whence 
$\theta_{ex} = 0$. For the single-spinon loop, the six local spinon-dimer 
exchange moves and the accompanying symmetric RK process contribute the 
same factor. However, the two additional RK processes of necessity have 
different shapes, which is critical in that the asymmetrical RK loop 
contributes a factor of $-1$. Thus we conclude that $\phi = \pi$ and 
hence $\theta_{s} = \pi$, meaning spinons are fermions. 

This result is due only to the rearrangement of dimers, i.e.~emerging 
fermionic statistics \cite{rht,riqf} are purely a consequence of the dimer 
background. Results for the 3D pyrochlore [Fig.~3(b)] are identical to those 
in 2D [Fig.~2]. The processes involve even numbers of local spinon moves and 
hexagonal RK loops symmetrical other than a single asymmetrical loop in the 
one-spinon process. 

\begin{figure}[t]
\includegraphics[width=3.8cm]{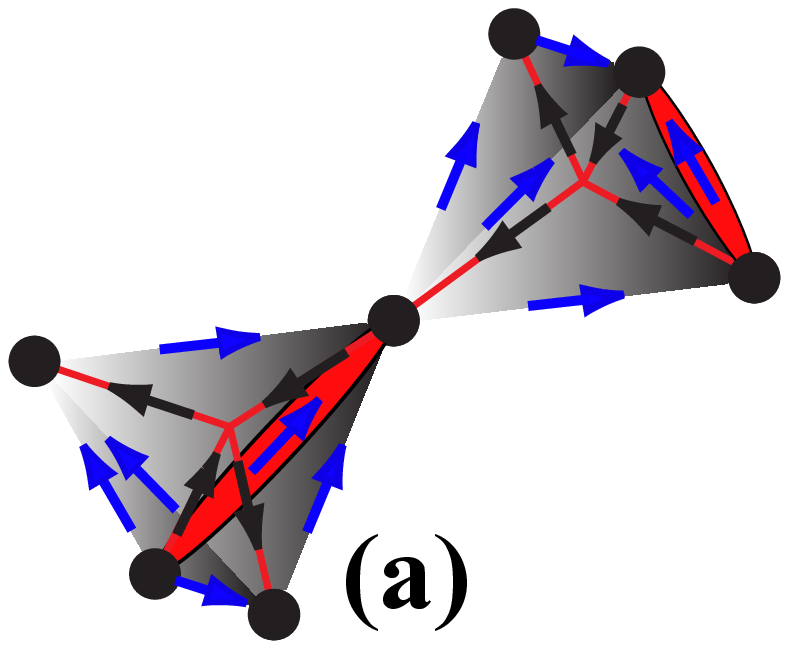}\hspace{1.0cm}\includegraphics[width=3.6cm]{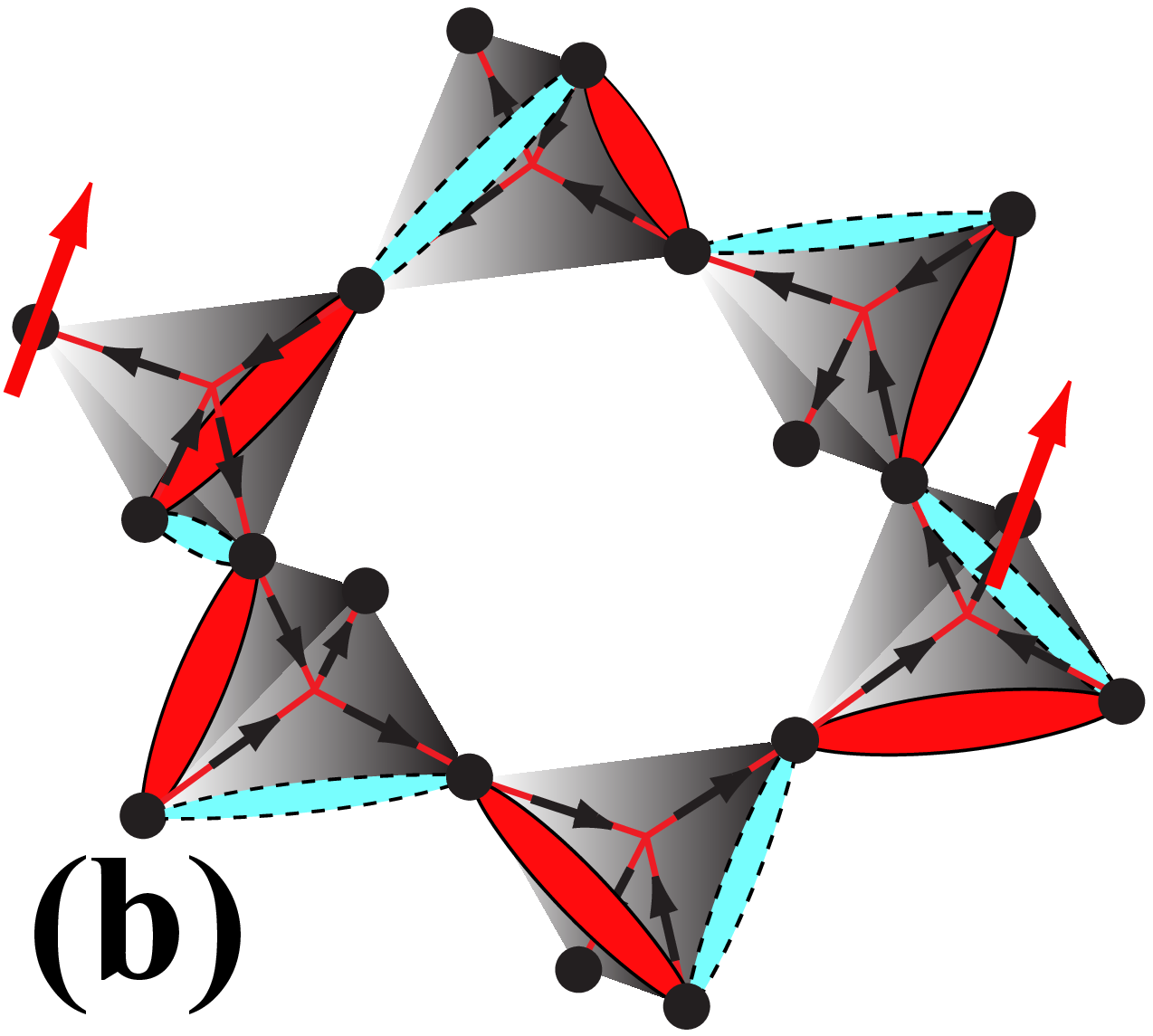}
\includegraphics[width=3.2cm]{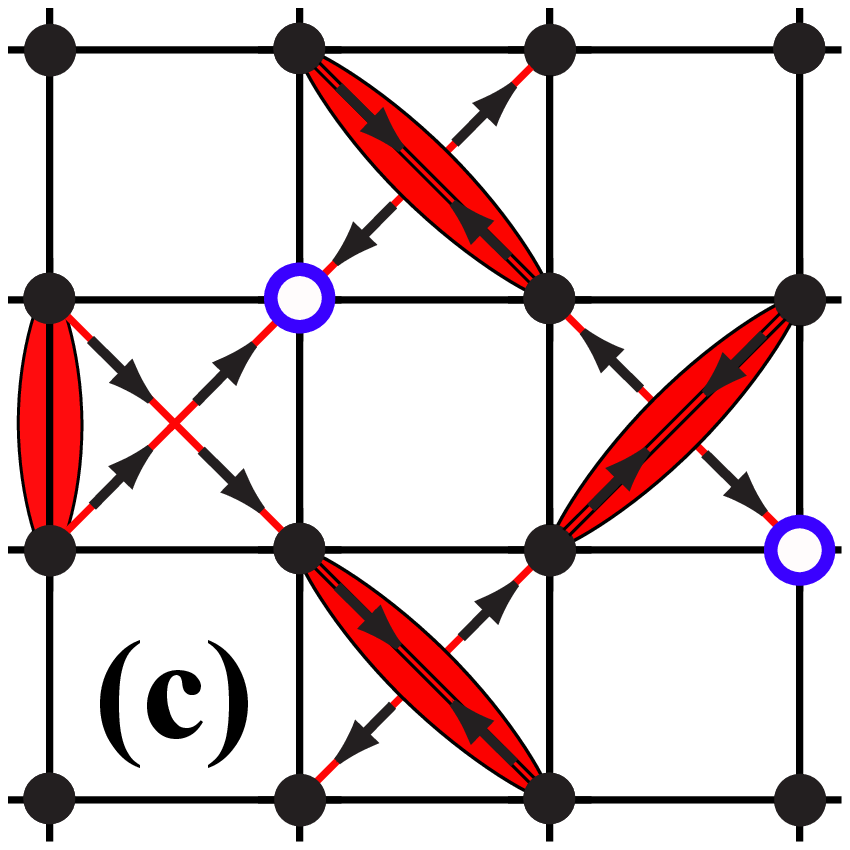}\hspace{1.2cm}\includegraphics[width=3.2cm]{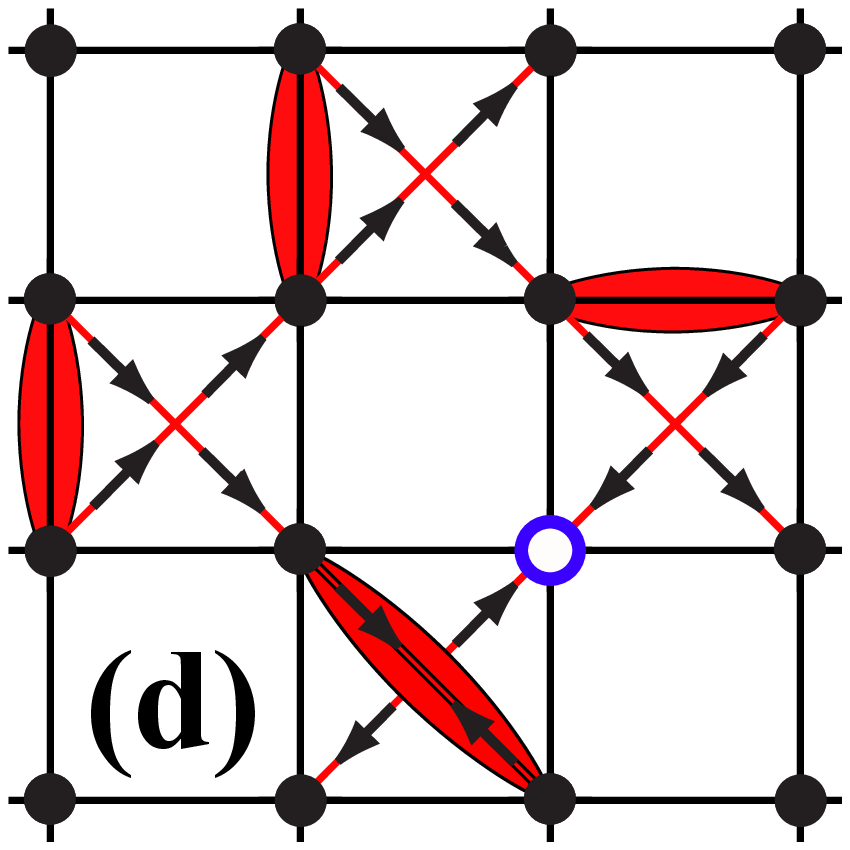}
\caption{(color online) Representation of (a) the sign structure adopted for 
the 3D pyrochlore; (b) two spinons beside a flippable loop allowing an exchange 
path around a single pyrochlore hexagon; (c) two holons on a flippable loop 
allowing an exchange path around a single checkerboard square; (d) a single 
holon on the same path.}
\label{fig3}
\end{figure}

To understand the origin of this result we consider an operator representation. 
Dimer singlets are bosonic objects created by an operator $d_{ij}^{\dag}
 = {\textstyle \frac{1}{\sqrt{2}}} (c_{i\uparrow}^\dag c_{j\downarrow}^\dag - 
c_{i\downarrow}^\dag c_{j\uparrow}^\dag)$, which is manifestly a composite of 
two fermionic electrons. At half-filling, there is no net charge motion 
and for processes within the ground manifold there is a meaningful sense 
in which $d_{ij}^{\dag} = {\textstyle \frac{1}{\sqrt{2}}}(f_{i\uparrow}^\dag 
f_{j\downarrow}^\dag - f_{i\downarrow}^\dag f_{j\uparrow}^\dag)$, where $f_{i\sigma}^\dag$ 
creates a spinon, an entity with only spin degrees of freedom. These spinon 
operators describe completely the changing spin correlations between sites 
during all processes changing the dimer coverings. As a consequence of 
perfect dimer singlet formation, spinons are fully spin-polarized, i.e.~they 
have zero entanglement with any dimer spins.

However, these same dimer rearrangement processes, specified above by the 
permutation operator, $P_{ij}$, are brought about by the real exchange of 
electrons, meaning that not only the spin but also the orbital degrees of 
freedom are exchanged. Because this process exchanges electrons described 
by $c_{i\sigma}^\dag$, it is not surprising that the spinons are fermionic. 
To state this result more explicitly, the many-body electronic wave function, 
$|\psi \rangle$, of the system changes sign under simultaneous exchange of 
all the orbital and spin degrees of freedom of any two electrons ($i 
\leftrightarrow j$, $\sigma_{i} \leftrightarrow \sigma_{j}$). Because 
the spinons correspond to two unentangled electrons of the same spin 
polarizations then by the fermionic character of the electrons, the 
exchange of spinons at sites $i$ and $j$ leads to a minus sign. Beyond 
the effective spinon description, and certainly at the scale of the Hubbard 
repulsion, $U$, appearing in the denominators of $J_1$ and $J_2$ (\ref{eht}), 
the preeminence of electronic statistics is clear; fermionic spinons are 
``emergent'' quasiparticles in the sense of the low-energy limit. 

\section{Hole Doping and Holon Statistics}

Next we consider the statistics of hole-like quasiparticles. Charge degrees 
of freedom arise on doping into the half-filled band and incur an energy 
penalty from the concomitant introduction of DTs. Spinon deconfinement 
causes immediate spin-charge separation \cite{rnn}, but ``holons'' also 
propagate in the dimer background through the kinetic term $- t 
\sum_{\langle ij \rangle, \sigma} c_{i\sigma}^{\dag} c_{j\sigma}$ of the Hubbard 
model. An important point not made in Ref.~\cite{rnn} is that the holon band 
width remains of order $t$ because, unlike the familiar square-lattice case, 
no local magnetic ordering acts to suppress hole motion.

Proceeding as for spinons, the action of $\sum_{\sigma} c_{k\sigma}^\dag c_{i\sigma}$ 
on state $|d_{ij} \rangle |0_k \rangle = {\textstyle \frac{1}{\sqrt{2}}} 
(|\! \uparrow \downarrow \! 0 \rangle - |\! \downarrow \uparrow \! 0 \rangle)$ 
yields ${\textstyle \frac{1}{\sqrt{2}}} (|0 \! \downarrow \uparrow \rangle$ 
$- |0 \! \uparrow \downarrow \rangle) \equiv |0_i \rangle |d_{kj} \rangle$, 
i.e.~there is an exact permutation of the hole state $|0_k \rangle$ with 
one end of the dimer, leading again to a sign-change along the loop. The 
full situation, depicted in Figs.~3(c) and 3(d) for two representative steps 
of the exchange and one-holon processes, is identical to that for spinons. 
The local quasiparticle moves give even numbers of $-1$ factors and the RK 
loops determine the statistics. We conclude that holons are fermions.

The most transparent way to understand holon statistics is to introduce 
them as a pair; we defer a discussion of the complexities in this process 
to Sec.~VIB. This pair replaces a single dimer, the operator expression of 
the process being $h_i^{\dag} h_j^{\dag} d_{ij}$, with $h_{i}^{\dag} = \sum_{\sigma} 
c_{i \sigma}$. Like spinons, both holons are entirely decoupled from the 
remaining dimers and hence their motion under all local dimer processes 
gives them fermionic statistics. A holon is simply the absence of an 
electron, is therefore fermionic, and may more accurately be termed a ``hole.'' 

In more detail, the hole at site $i$ may be regarded as an electron that was 
initially in a localized orbital state at $i$ but subsequently ejected from 
the system, retaining in the process its original spin degrees of freedom 
and creating a state charged relative to its background. When two such 
excited electrons, associated with initially localized states at sites $i$ 
and $j$, are ejected from the many-body wave function, $|\psi \rangle$, 
exchanging their orbital (and charge) degrees of freedom leads to $h_i^\dag 
h_j^\dag P_{ij} |\psi \rangle = - h_i^\dag h_j^\dag |\psi \rangle$, i.e.~hole 
permutation is fermionic. We observe in addition that the action of 
$h_{i}^{\dag}$ on the dimer state $|d_{ij} \rangle$ leaves ${\textstyle 
\frac{1}{\sqrt{2}}} (c_{j\downarrow}^{\dag} - c_{j\uparrow}^{\dag})$, which specifies 
a spinon state, of no relative charge and perfect spin polarization along 
$- {\hat x}$. We stress that there is no sense in which spinons and holons 
are fractionalized electrons; instead they are fractionalized dimers, these 
being the fundamental objects of the undoped ground state.

The key to this result is that both spinon and holon quasiparticles are fully 
specified by electron operators. Thus the underlying reason for their fermionic 
statistics is explicit. This is not a consequence of any special model such as 
the pyrochlore QSL, which we use here to make our statements rigorous. Quite 
generally, in any electronic dimer state, however complex, both holon 
and spinon dynamics arise ultimately from fermionic electron motion. This 
result is completely universal, and it is clear from the definition of the 
quasiparticles in terms of electronic operators that it overrides all details 
such as longer-ranged dimers and local processes, overcomplete basis sets, or 
excitations (``visons'') connecting different topological sectors. 

More subtle is the question of whether quasiparticle statistics can be 
dictated by the signs of the loops, meaning the overlap matrix elements 
between different dimer configurations in the ground manifold. In 
Ref.~\cite{rlrcpp} it was shown that the statistics of holons doped in 2D 
QDMs can be exchanged from fermionic to bosonic by exchanging the sign of 
the quantum dimer kinetic term, which is a matrix element. This result, 
equivalent to the discussion of flux attachment \cite{rk,rrc}, was later 
generalized to include the signs of all loop processes in the system 
\cite{rlropp}. Although the sign of the interaction term is fixed in our 
electronic model, it is possible to change the phase relation between 
the different dimer configurations appearing in the linear superposition of 
coverings, $|\psi \rangle = \sum_a c_a |\psi_a \rangle$, making up the ground 
state \cite{rnn}. However, unlike the situation in 2D, where even in 
frustrated models it may be possible to exchange the relative signs of all 
pairs of dimer coverings, $|\psi_a \rangle$, in the ground state differing 
by short loops \cite{rnbnt}, it is manifestly impossible to achieve this for 
all the highly overlapping short loops arising in the 3D system \cite{rnn}. 
Thus we must conclude that quasiparticle statistics are not arbitrary in 3D; 
if the electronic state has a valid dimer description, then the quasiparticles 
have explicit representations in terms of electron operators and they have 
fermionic statistics. The only possible exceptions arise in 2D systems 
allowing the attachment of statistical flux, which has an alternative 
interpretation in terms of a loop sign ambiguity. 

\section{Fermions, Strings, and Gauge Fields}

Emergent fermionic quasiparticles, their gauge symmetry, and the extended 
entities (``strings'') they form are the three fundamental concepts introduced 
in Ref.~\cite{rlw}. The general case of arbitrary quasiparticle number reflects 
the gauge symmetries, or local conservation laws, of the system. The only 
strict local conservation law, the zero-divergence condition, is one dimer per 
tetrahedron. In the presence of quasiparticles and DTs, one has $n_{di} + n_{DTi} 
 = 1$ for the dimer and DT numbers on every tetrahedron $i$. Each DT introduces 
two free quasiparticles, restricted only by a global conservation law, 
$\Sigma_i [(n_{si} + n_{hi}) - 2 n_{DTi}] = n_s + n_h - 2 n_{DT} = 0$. This 
expression corresponds directly to the ``effective charges'' of spinons 
($-1$), holons ($-1$), and DTs ($+2$) \cite{rnbnt,rnn}. The holon number is 
the dopant concentration, $\Sigma_i 2 n_{hi}/N = 2 n_h/N = x$, which specifies 
the ``charge sector'' of the system, another global constraint. Encoded as a 
gauge principle, the sole local constraint corresponds to a U(1) gauge 
theory \cite{rhfb}, and hence in the pyrochlore QSL, even with doping, 
only one U(1) symmetry emerges from the local physics of the dimers and 
quasiparticles. 

\begin{figure}[t]
\includegraphics[width=4.1cm]{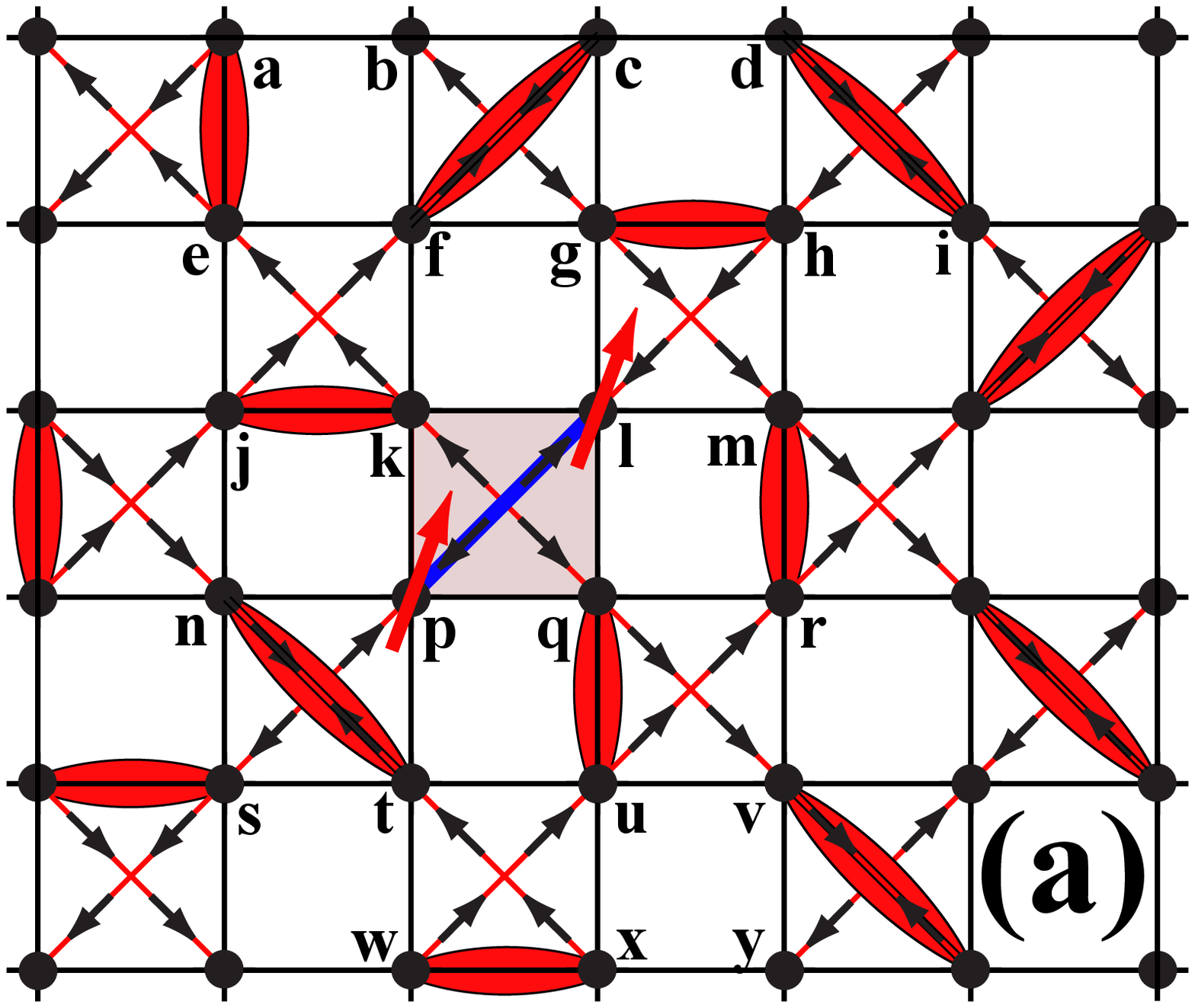}\hspace{0.2cm}\includegraphics[width=4.1cm]{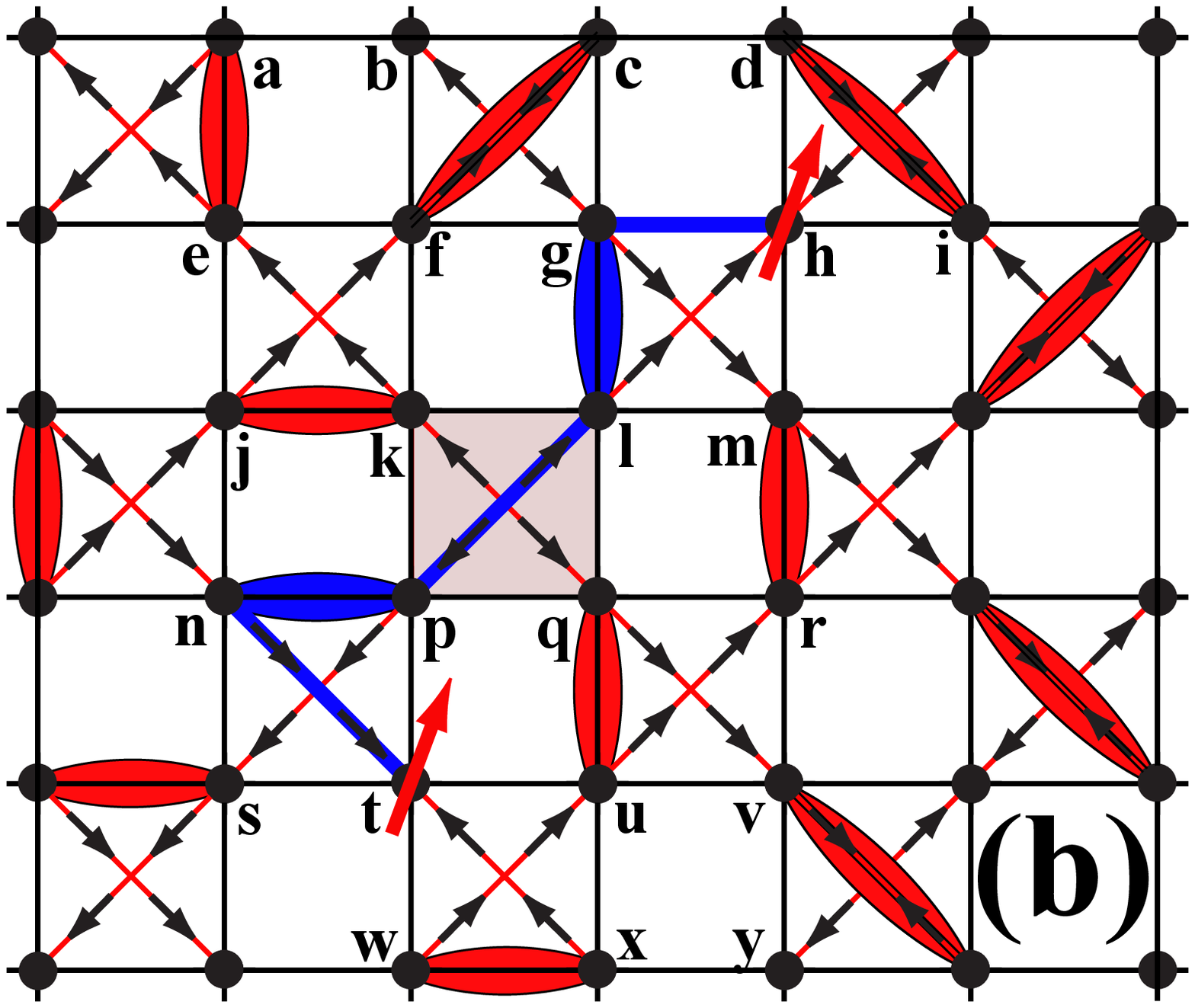}
\includegraphics[width=4.1cm]{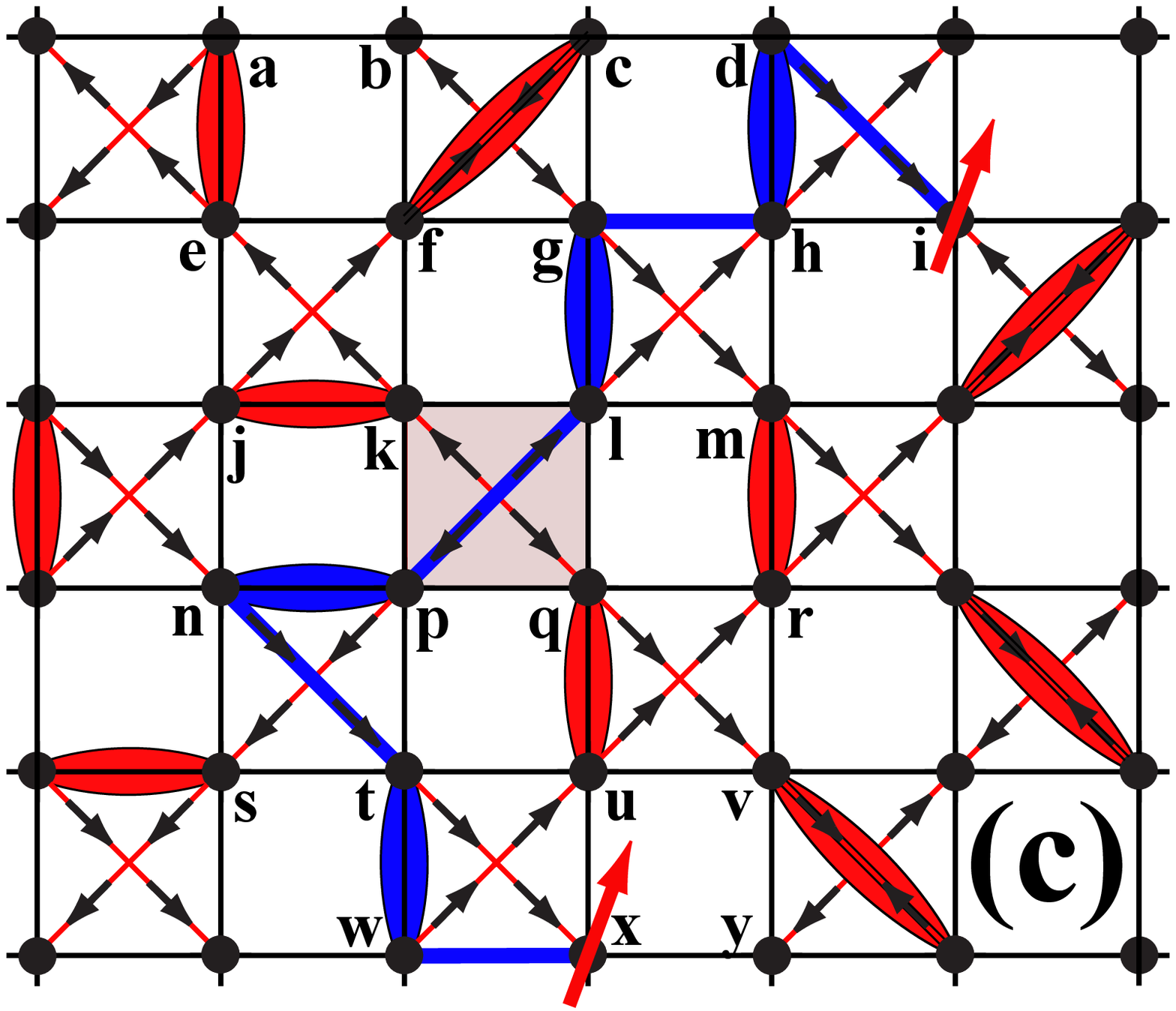}\hspace{0.2cm}\includegraphics[width=4.0cm]{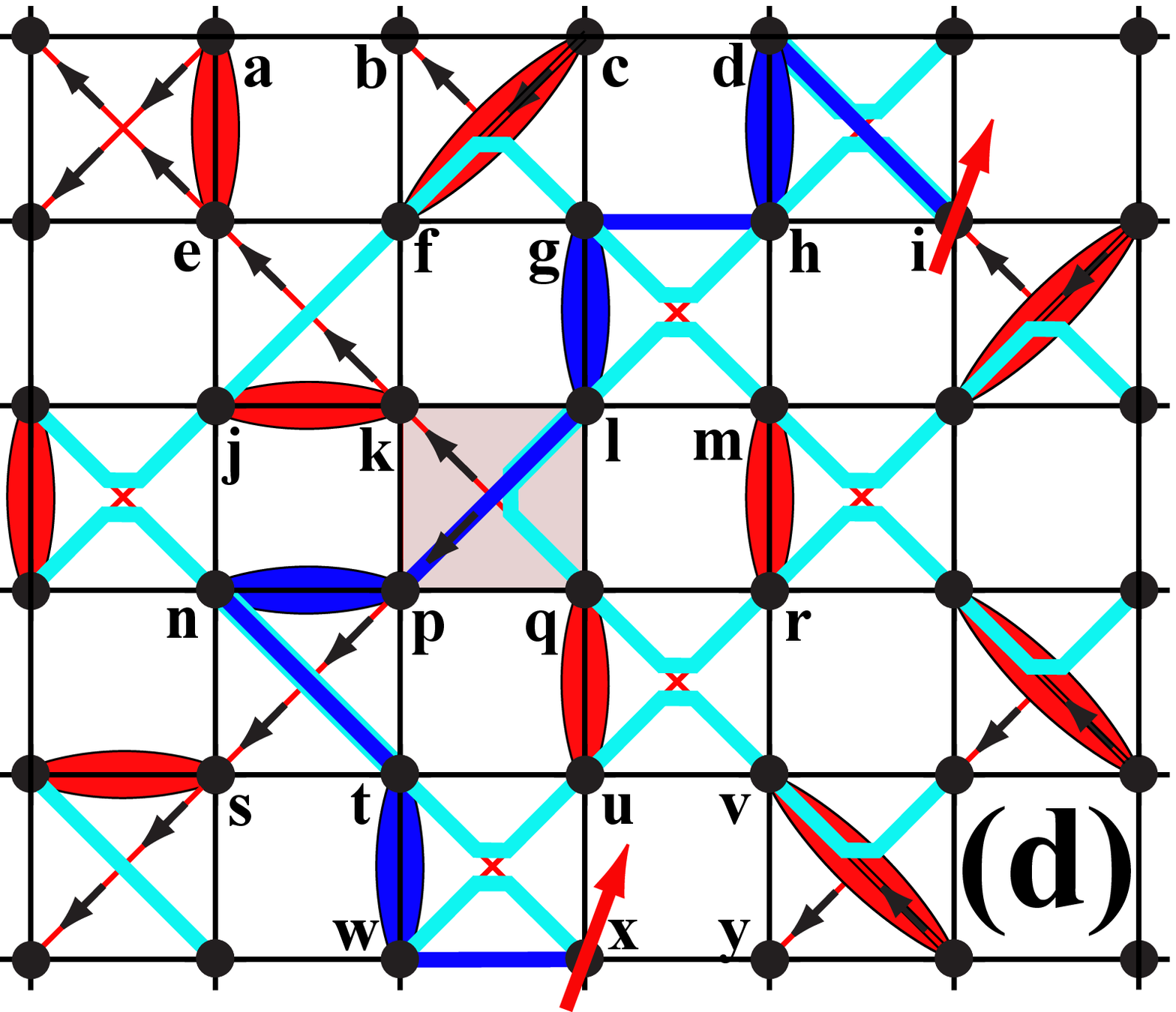}
\caption{(color online) Representation of (a) a single defect tetrahedron with 
two quasiparticles on the DT, (b) the first step for each quasiparticle away 
from the DT, and (c) the second, leaving a string. (d) Line representation and 
gauge string shown for the same quasiparticle propagation process.}
\label{fig4}
\end{figure}

To make this gauge symmetry completely rigorous, and its connection 
to ``strings'' \cite{rlw} more transparent, we express the system as 
a lattice gauge theory following Ref.~\cite{rnna}. The presence 
or absence of a dimer connecting any two sites $i$ and $j$ in the same 
tetrahedron is denoted by the states $\sigma^z_{ij} = \pm 1$, with 
corresponding singlet creation and destruction operators $\sigma^\pm_{ij}
 = {\textstyle \frac12} (\sigma^x_{ij} \pm i \sigma^y_{ij})$. Any basis state 
may be specified only by $\sigma^+$ operators for all pairs of sites in the 
dimer covering $a$, $|\psi_a \rangle = \Pi_a \sigma^+_{\langle ij \rangle \in a} 
|0 \rangle$. Any loop process between two such states is specified by a 
sequence $\sigma^+_{ab} \sigma^-_{bc} \sigma^+_{cd} \sigma^-_{de} \sigma^+_{ef} 
\sigma^-_{fg} \sigma^+_{gh} \sigma^-_{ha}$ connecting the dimers along alternate 
bonds, with two operators in each tetrahedron. A quasiparticle is specified 
by $... \sigma^-_{ab} \sigma^+_{bc} \sigma^-_{cd} {\overline q_d} \sigma^-_{de} 
\sigma^+_{ef} \sigma^-_{fg} ...$, where the two $\sigma^-$ operators appearing 
in sequence encode its presence on site $d$ (${\overline q_d}$ is required 
to specify a spinon or holon). 

Creating a DT in the basis wave function $|\psi_a \rangle$ replaces 
$\sigma^+_{lp}$ by ${\overline q}_l \sigma^-_{lp} {\overline q}_p$ in Fig.~4(a). 
The local processes causing the two quasiparticles to move away from the DT 
are ${\overline q}_h \sigma^-_{gh} \sigma^+_{gl} \sigma^-_{lp} \sigma^+_{pn} 
\sigma^-_{nt} {\overline q}_t$ [Fig.~4(b)] and then ${\overline q}_i 
\sigma^-_{di} \sigma^+_{dh} \sigma^-_{gh} \sigma^+_{gl} \sigma^-_{lp} \sigma^+_{pn} 
\sigma^-_{nt} \sigma^+_{tw} \sigma^-_{wx} {\overline q}_x$ [Fig.~4(c)]. 
Schematically, the propagation of the quasiparticles develops an extended 
object, a string. Mathematically, this exact spin representation is 
transformed to a lattice gauge theory by $\sigma^\pm = e^{\pm i A_{ij}}/ \! 
\sqrt{2}$ with a real, compact phase $A_{ij}$, which is canonically 
conjugate to $\sigma^z_{ij}$, as the lattice gauge field \cite{rnna}. 
Destroying the dimer on a tetrahedron is described by $A_{ij} \longrightarrow
 - A_{ij} + \theta_i + \theta_j$, where $\theta_i$ is a U(1) phase degree of 
freedom carried by the quasiparticle ${\overline q}_i$ created on site $i$. 
As the quasiparticles propagate, lengthening their string (Fig.~4), this 
U(1) phase information is retained by the moving particles and its memory, 
effectively the fingerprint of the missing dimer, is preserved in the phase 
$-A_{ij}$ on the DT pinning the string. 

The lattice gauge theory of Ref.~\cite{rnna} is an exact representation 
of the ground manifold of dimer coverings. In the presence of quasiparticles, 
introduced as above, its extension is a matter-coupled gauge theory of the 
form ${\cal L} = {\overline q}_i U_{ij} q_j + {\rm c.c.}$. This is a 
``minimal-coupling'' gauge theory where the fermionic quasiparticles appear 
as source terms of the gauge field and the Lagrangean is invariant for 
all lattice sites under the U(1) gauge symmetry $U_{ij} \rightarrow \eta_i^* 
U_{ij} \eta_j$, $q_i \rightarrow q_i \eta_i^*$, with $\eta_i \equiv e^{i \theta_i}$. 

However, it is essential to note here that the strings, or gauge fields, 
carry no energy in the pyrochlore QSL. As a consequence of the massive 
degeneracies of the ground and first-excited manifolds, all quasiparticle 
locations are entirely equivalent in energy and so spinon and holon motion 
is deconfined. The role of strings is to mediate the phase relationship 
connecting the quasiparticles, via their point of origin at a DT, but this 
U(1) phase does not correspond to a physical observable.

\section{Strings and Physics}

\subsection{Strings and Lines}

The strings (of $\sigma$ operators) in Fig.~4 are a transparent way to 
consider quasiparticle propagation. There are two $\sigma$ operators on 
every tetrahedron of the string except for the DT, where there is only 
one. These strings are not the same as the lines used in the line 
representation \cite{rnbnt}, as shown in Fig.~4(d). Strings encode the 
U(1) phase information of the spinon/holon pair and the DT, i.e.~they 
are specific to defect formation and phase conservation laws. Lines 
include all dimers and encode their local conservation law. The maximally 
flippable submanifold is the set of all states with an average of 1 line 
per tetrahedron [Fig.~1(a)], which is the most degenerate sector. 

However, spinons and holons are both the endpoints of both lines and strings 
(Fig.~4). There is only one type (``flavor'') of line and one type of string, 
which may end with either type of quasiparticle, and local processes allow 
quasiparticles to exchange the lines/strings of which they are endpoints. 
Both lines and strings are pinned to DTs. Lines do not cross, whereas 
strings from different pairs of quasiparticles may do so. Both lines and 
strings are useful graphical representations of the fact that both species 
of quasiparticle, although energetically deconfined (free to move anywhere 
at zero energy cost), do retain a memory of their origin. As one consequence 
of this, the two quasiparticles from the same DT cannot repair each other's 
tracks, whereas quasiparticles of different origins may do so in part. 

\subsection{Strings and Local Quasiparticles}

In Secs.~III and IV, we treated the quasiparticles as local objects in 
order to deduce their statistics, and in fact related them directly to 
local electron operators. However, in Sec.~V we discussed their connection 
to strings, which suggests the relevance of extended objects. Indeed, it 
was pointed out in Ref.~\cite{rlw} that emergent fermions are non-local 
objects and should always be considered in pairs, whence our consideration 
of doped holon pairs in Sec.~IV. 

There are two issues to discuss here. First, on the question of whether 
quasiparticles are extended objects or not, the answer is already clear 
from Sec.~V. Quasiparticles are deconfined objects and their gauge strings 
have no energy, acting only as a book-keeping device for their phase. This 
defect-related phase variable is not a physical observable and does not 
appear in the quasiparticle statistics. DTs also have no dynamics and make 
no contributions to the statistics. Thus the quasiparticles can be treated 
as local objects. The only conceptual point here is the rather trivial one 
that it does not make sense to attempt to define the statistics using only 
one quasiparticle, and that this must be done by considering pairs. 

The second issue is how to dope holes into a dimer-based system. Clearly, 
quasiparticles always replace dimers and thus by definition are introduced 
in pairs. Spinons are created automatically in pairs by the excitation of a 
single dimer; this process also creates one DT, with which the spinons 
maintain a fixed phase relation, but none of their physical properties are 
affected by it. By contrast, in the case of holons it is necessary to define 
a valid creation process, because in principle a single holon can be inserted 
by the elimination of a single electron from the ground manifold. However, 
this cannot be the complete process in a dimer-based system, such as the 
pyrochlore QSL, because the other electron from the destroyed dimer remains
present; from Sec.~IV, it will be a fully spin-polarized electron, i.e.~a 
spinon. To avoid spinon interference, holon statistics are best considered 
by introducing them in pairs that replace a single dimer, also creating one 
DT, as in Sec.~IV.

When a single hole is introduced in the system, its lonely ex-partner 
spinon may form a new singlet, the new unpaired spin forms another new 
singlet, and so on, in a sequence of reconstructed spin correlations 
equivalent to the propagation of the spinon away from the doped holon. 
Despite the appearance of an extended object, as above this string has no 
physical meaning. For the purposes of deducing exchange statistics, the 
quasiparticles in the exchange process originate in general from different 
DTs; whether the holons of an exchanged pair were introduced singly or 
pair-wise is not relevant, as long as single-hole doping is taken to replace 
each bosonic dimer by one holon and one spinon \cite{rnn}. This satisfies 
the constraints that two quasiparticles are always accompanied by one DT and 
that the total number of quasiparticles is never odd. The possibility that 
the total number of spinons, or of holons, is individually odd has no effect 
on exchange considerations in the assumed limit of dilute quasiparticles. We 
reiterate that the process of creating quasiparticles is a fractionalization 
of dimers and our contribution here is to prove that both fractions are 
fermionic. 

\section{Summary}

We have considered the nature of quasiparticles in the pyrochlore quantum 
spin liquid. Both spinons and holons have fermionic statistics. These 
properties are conferred entirely by the dimers of the highly degenerate 
basis manifold and are completely universal for all electronic dimer states. 
Spinons and holons are linked by strings, which correspond to a gauge field 
whose origin lies in the conservation of dimer number. 

The lattice gauge theory of the pyrochlore QSL is not only an effective 
description but an exact recasting of the fact that all dimer states and 
loop processes are known exactly. The local ($d = 0$) U(1) gauge field 
expresses the local constraint and is connected with the emergent pyrochlore 
photon \cite{rhfb}, a mode whose gaplessness is a consequence of the 
degeneracy of the ground manifold and whose linearly dispersive character 
is determined by the nature of the excited manifolds. Off the Klein 
point, additional planar ($d = 2$) gauge terms specify the new ground 
manifold \cite{rnn}. In the presence of spinons or holons, one has a 
matter-coupled gauge theory with minimal coupling (${\overline q}_i U_{ij} 
q_j \; + \; {\rm c.c.}$) of fermionic quasiparticle matter to a dimer-mediated 
bosonic field and U(1) gauge symmetry. 

Concerning possible superconductivity in the doped pyrochlore QSL, minimal 
renormalization of the hopping $t$ means significant kinetic energy may be 
gained by holon motion. A small concentration of holes doped into the 
insulating half-filled system will create a small Fermi surface. Given the 
attractive but weak potential caused by the immobile DTs \cite{rnn}, it 
appears likely that holons, being fermionic, will pair and then superconduct 
at suitably low temperatures, presenting a specific example of RVB-type 
superconductivity \cite{rhtsc1} at low doping. 

We close with a brief discussion of the three key features exhibited by
the pyrochlore QSL. The fundamental property of the pyrochlore geometry is 
its four-site unit, which allows each to contain one dimer in the ground 
manifold. This dimer-based structure establishes the local constraint 
determining the quasiparticle statistics. The constraint also links the 
creation of propagating quasiparticles to the two sites of their DT. 
Schematically, the composite operation $e^{i\theta_i} {\overline q}_i 
e^{i\theta_j} {\overline q}_j e^{-i\theta_{ij}} d_{ij}$ has a phase degree of 
freedom restricted by $\theta_i + \theta_j - \theta_{ij} = 0$, where $\theta_i$ 
is a U(1) quasiparticle phase and $\theta_{ij} = 2 A_{ij}$ is given by the 
lattice gauge field. This link remains present as a gauge string and the 
memory of the missing dimer (violation of the constraint) is preserved in 
the wave function through $A_{ij}$. Finally, the fermionic nature of both 
spinons and holons is no big mystery. Both are the physical quasiparticles of 
the starting Hubbard model. There is no sense in which one is fractionalizing 
an electron, only pairs of electrons (dimers). The resulting fractions are 
fermions, the spinon having the essential characteristics of the electron 
and the holon of a missing electron.

\acknowledgments

We thank C. D. Batista and X.-G. Wen for helpful discussions. This work was 
supported by the NBRP of China under Grant No.~2012CB921704, by the NSF of 
China under Grant No.~11174365, and by the NSF under Grants CMMT 1106293 and 
PHY11-25915.


\begin{thebibliography}{99}
\bibitem{cpc} Because we continue to rely on the checkerboard geometry for 
explanatory purposes in Figs.~2--4, it is essential to clarify that, although 
quasiparticle confinement differs from the pyrochlore for dimensional reasons 
\protect{\cite{rnn}}, the physics of local processes is identical and of loop 
processes is exactly analogous between the vacant squares of the checkerboard 
and the hexagons of the pyrochlore.
\end{thebibliography}
\end{document}